\documentclass[aps,preprint,floatfix]{revtex4}
\usepackage{amsmath}
\usepackage{latexsym}
\usepackage{float}
\usepackage{amssymb}
\usepackage{graphicx}
\newcommand{\dummy}

\begin{document}

\title{Kinetics of Phase Separation in Thin Films: Simulations for
the Diffusive Case}

\author{Subir K. Das$^{\text{1}}$, Sanjay Puri$^{\text{2}}$, J\"urgen
Horbach$^{\text{1}}$ and K. Binder$^{\text{1}}$}
\affiliation{$^{\text{1}}$\textit{Institut f\"{u}r Physik, Johannes
Gutenberg-Universit\"{a}t} \\ \textit{D-55099 Mainz, Staudinger Weg 7,
Germany} \\
$^{\text{2}}$\textit{School of Physical Sciences,
Jawaharlal Nehru University, New Delhi-110067, India}}

\begin{abstract}

We study the diffusion-driven kinetics of phase separation of a
symmetric binary mixture (AB), confined in a thin-film geometry
between two parallel walls. We consider cases where (a) both walls
preferentially attract the same component (A), and (b) one wall
attracts A and the other wall attracts B (with the same strength).
We focus on the interplay of phase separation and wetting at the
walls, which is referred
to as {\it surface-directed spinodal decomposition} (SDSD).
The formation of SDSD waves at the two surfaces, with
wave-vectors oriented perpendicular to them, often results in a
metastable layered state (also referred to as ``stratified
morphology''). This state is reminiscent of the situation where
the thin film is still in the one-phase region but the surfaces are
completely wet, and hence coated with thick wetting layers. This
metastable state decays by spinodal fluctuations and crosses over
to an asymptotic growth regime characterized by the lateral
coarsening of pancake-like domains. These pancakes may or may
not be coated by precursors of wetting layers. We use Langevin
simulations to study this crossover and the growth kinetics in
the asymptotic coarsening regime.

\end{abstract}

\maketitle

\newpage

\section{INTRODUCTION}

Consider a binary mixture (AB) with a miscibility gap, such
that phase separation into A-rich and B-rich phases occurs below
the critical temperature $T_c$. If this mixture is quenched
from a homogeneous state in the one-phase region into the two-phase
region below the critical point, phase separation proceeds by the
emergence and growth of regions enriched in either component. In
the bulk, this process of {\it spinodal decomposition} or
{\it domain growth} has been intensively studied \cite{1,2,3,4,5}.

In recent years, the effect of surfaces on this behavior
has become a focus of research \cite{6,7,8,9,10,11,12,13}. It is
often the case that a surface (S)
has a preferential attraction for one of the
components (say, A) of the mixture. In the one-phase
region of the mixture, this attraction leads to the formation of
surface enrichment layers of the preferred component. The
thickness of these layers (which is comparable to the
correlation length $\xi_b$ of concentration fluctuations in the
bulk) becomes long-ranged at the critical point of the mixture
\cite{14,15}. In addition, if the composition of the bulk
mixture coincides with the B-rich branch of
the coexistence curve describing phase separation in
equilibrium, the surface is coated with an A-rich
wetting layer for temperatures below the wetting transition
temperature $T_w$ \cite{16,17,18,19,20,21,22}. Of course, wetting
layers of mesoscopic thickness (in the idealized case, the
thickness of the wetting layer diverges when complete wetting
occurs \cite{19,20,21,22}) can only occur in
macroscopic systems. Typically, theoretical approaches
consider the idealized case of a
semi-infinite geometry \cite{14,15,16,17,18,19,20,21}.

\begin{figure}[htb]
\includegraphics[width=75mm]{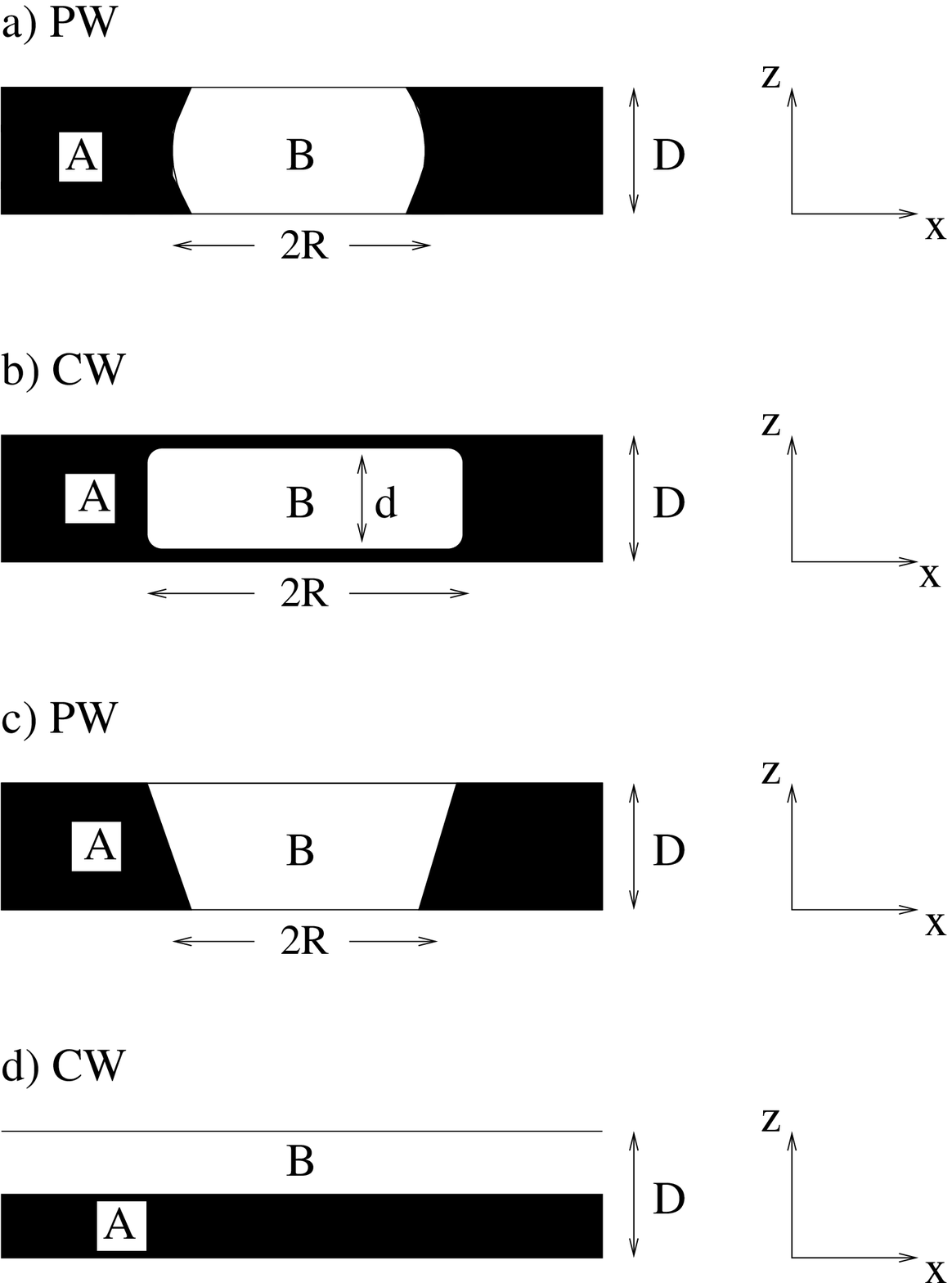}
\caption{\label{fig1}Schematic description of phase-separated
states in thin films of volume $L \times L \times D$. We show
cross-sections of the film in the $(xz)$-plane. Cases (a)-(b)
refer to thin films where both surfaces prefer A, and the
surface potential is symmetric, $V(D-z)=V(z)$. Cases (c)-(d)
refer to thin films where the lower surface prefers the A-rich phase
and the upper surface prefers the B-rich phase. The corresponding
surface potential is antisymmetric, $V(D-z)=-V(z)$. The A-rich
domains are marked black, and the B-rich domains are unmarked.
For both symmetric and antisymmetric films,
partially wet (PW) and completely wet (CW)
morphologies emerge in the limit $D \rightarrow \infty$.
Further types of phase-separated states exist in thin films at
off-critical composition \cite{34,35,36}, but we will not
consider these here. Note that the thickness $d$ of
the encapsulated B-rich domain in (b) differs from $D$ only by
corrections which increase slower than linearly with $D$.}
\end{figure}

In a thin film of thickness $D$, finite-size effects significantly
constrain the growth of wetting layers, e.g., for short-range
forces between the walls and the A-atoms, the equilibrium
thickness of a wetting layer is O($\ln D$) \cite{22,23,24,25}.
Further, in a thin-film geometry, the interplay
between surface effects and bulk phase separation leads to a
distortion of the phase diagram describing lateral phase
separation parallel to the walls, analogous to the phenomenon of
capillary condensation \cite{22,24,25,26,27,28,29}. Moreover,
the critical behavior changes its character from three-dimensional
to two-dimensional \cite{30}. The interplay of wetting and phase
separation in a thin film results in a rich phase
behavior, and the equilibrium phase diagrams of thin
films have been investigated for a range of possible
surface forces \cite{22,29,31,32,33,34,35,36}. In the
low-temperature region, relevant for deep quenches to a
temperature $T<T_w$, the typical state in a thin film with
symmetric walls (i.e., both attract A with the same strength)
is a laterally-segregated state with a
B-rich pancake-like domain. This domain is circular
with a macroscopic radius $R$, which is of the same order as the
macroscopic linear dimension $L$ parallel to the walls -- see
Fig.~\ref{fig1}(a). For a volume fraction
$\phi_B$ of B-atoms lying between the two branches of the
coexistence curve $\phi_{B, {\rm coex}}^{(1)} (D)$, $\phi_{B, {\rm
coex}}^{(2)} (D)$ of the thin film, the lever rule requires
\begin{equation} \label{eq1}
\phi_B=(1-x)\phi^{(1)}_{B, {\rm coex}} (D) + x \phi^{(2)}_{B, {\rm
coex}} (D) \quad ,
\end{equation}
where $x$ is the volume fraction of the B-rich phase. For
a film of volume $L^2D$ (Fig.~\ref{fig1}) with periodic boundary
conditions in the directions parallel to the walls, we have
$x=\pi R^2/L^2$. Assuming a strongly segregated state where
$\phi^{(1)}_{B, {\rm coex}} (D) \simeq 0$, $\phi^{(2)}_{B, {\rm
coex}} (D) \simeq 1$, we obtain $R \simeq L \sqrt{\phi_B/\pi}$
for the radius of the B-rich domain.

In this simple argument, we have assumed that the
pancake is cylindrical and its surface is
perpendicular to the walls. In fact, the interface is curved
because (in the $D \rightarrow \infty$ limit) the A-B interface
meets the wall at a contact angle $\theta$, given by Young's
equation \cite{37}:
\begin{equation} \label{eq2}
\sigma \cos \theta= \gamma_B - \gamma_A \quad .
\end{equation}
Here, $\sigma$ is the interfacial tension between the A-rich and
B-rich phases; and $\gamma_A$, $\gamma_B$ are the surface
tensions between the A-rich and B-rich phases and the wall,
respectively. The surface is partially wet (PW) when
$\sigma > \gamma_B -\gamma_A$ \cite{16,17,18,19,20,21,22}.
When $T \rightarrow T_w^-$, $\theta \rightarrow 0$,
and for $T>T_w$ one has $\sigma < \gamma_B - \gamma_A$. In this
complete wetting (CW) situation shown in Fig.~\ref{fig1}(b),
there is no direct contact between the
B-rich phase and the wall -- the B-rich pancake
is encapsulated by A-rich wetting layers for $D \rightarrow
\infty$. (For finite $D$, only precursors of wetting layers are
possible.) The cross-sections of these states [Figs.~\ref{fig1}(a)-(b)]
are reminiscent of the states considered by Liu et al. \cite{38}
in the context of phase separation in cylindrical pores. Liu et al.
made a distinction between elongated plugs
in the PW case and capsules in the CW
case, respectively. Note, however, that the lateral size
of these domains in equilibrium for small $D$ and $L \rightarrow
\infty$ does not diverge with $L$, unlike the present
$d=2$ cases. Rather the lateral size $\ell \sim
\exp [\sigma \pi D^2/(4 k_BT)]$, where $D$ is the pore diameter.
We also note that, in the CW case the thickness of
the B-rich domain differs from $D$ only by a small amount, that
will be estimated below.

It is also interesting to consider equilibrium morphologies in
antisymmetric films, i.e., one wall attracts A and the other
attracts B with equal strength. In Fig.~\ref{fig1}(c), we show
a schematic of the PW state in an antisymmetric film. This state
also resembles a pancake, except that the cross-section is
trapezoidal in this case due to the difference in contact angles
at the two walls. In Fig.~\ref{fig1}(d), we show a schematic of
the CW state in an antisymmetric film. In this state, there is
a single A-B interface which is parallel to the walls.

In this paper, we are interested in the kinetic processes
which lead to the formation of these structures subsequent
to a deep quench from the high-temperature disordered state. In
these {\it surface-directed spinodal decomposition} (SDSD)
processes, one typically observes first the formation of a layered
structure, consisting of an A-rich layer followed by a depletion
layer in A, etc. This layered profile
propagates into the bulk, but at a later stage it may break up
into a laterally inhomogeneous structure. The kinetics of such
growth phenomena has important technological applications,
including the fabrication of nanoscale patterns and layered
structures. However, despite much effort (for
reviews, see \cite{9,10,11,12,13}) these phenomena are still
incompletely understood.

In the present paper, we will study the dynamics of these
structure-formation processes for the case where hydrodynamic
effects can be disregarded, so that the kinetics is purely diffusive.
Therefore, our results will be applicable to solid mixtures, and the
early stages of phase separation in polymer and fluid mixtures.
(In a subsequent paper, we will elucidate the role of hydrodynamic
effects in these problems.) We shall consider the cases of both
symmetric and antisymmetric films. This paper is organized
as follows: In Sec. II, we
describe our dynamical model for segregation in confined
geometries. In Sec. III, we focus on the kinetics of
phase separation in a symmetric film, while Sec. IV studies
segregation kinetics in an antisymmetric film. We conclude this
paper with a summary and discussion in Sec. V.

\section{Dynamical Model for Segregation in Films}

The model for phase separation at surfaces has been developed
previously by Puri and Binder \cite{7}. In dimensionless units
\cite{10,13}, the free energy functional for an unstable binary
mixture in a film extending from $z=0$ to $z=D$ is \cite{39}
\begin{eqnarray} \label{eq3}
F[\psi] &\simeq& \int d \vec{r} \left[ -\frac{\psi^2}{2} +
\frac{\psi^4}{4} + \frac{1}{4} (\vec{\nabla} \psi)^2 + V(z)
\psi\right] + \nonumber \\
&& \int\limits_{S_1} d \vec{\rho} \left\{- \frac{g}{2}
[\psi(\vec{\rho},0)]^2-h_1 \psi (\vec{\rho}, 0) - \gamma \psi
(\vec{\rho}, 0) \frac{\partial \psi}{\partial z} \bigg| _{_{z=0}}
\right\} + \nonumber \\
&& \int\limits_{S_2} d \vec{\rho} \left\{- \frac{g}{2}
\left[\psi(\vec{\rho}, D) \right]^2 - h_2 \psi (\vec{\rho}, D) +
\gamma \psi(\vec{\rho}, D)
\frac{\partial \psi}{\partial z} \bigg| _{_{z=D}} \right\} \nonumber \\
&\equiv& F_b + F_{S_1} + F_{S_2} \quad .
\end{eqnarray}
Here, $\psi(\vec{r})$ is the order parameter which is proportional
to the density difference between the two species, $\psi(\vec{r})
\propto \rho_A(\vec{r}) - \rho _B (\vec{r})$. It is normalized
such that the coexisting A-rich and B-rich bulk phases for $T<T_c$
correspond to $\psi = \pm 1$, respectively.
The first term on the RHS of Eq.~(\ref{eq3}) is
the bulk free energy $F_b$, with $V(z)$ being the $z$-dependent
potential due to the surfaces $S_1$ and $S_2$.
In our subsequent discussion, we will consider power-law
potentials: $V(z)=-V_0[(z+1)^{-n} \pm (D+1-z)^{-n}]$, where
the $+$ and $-$ signs denote symmetric and antisymmetric films,
respectively. Such potentials are common in the context of
surface-molecule interactions, e.g., $n = \kappa - d$ with
$\kappa = 6$ and 7 corresponds to cases with non-retarded
and retarded van der Waals' interactions, respectively \cite{dlp61}.
The potentials are taken to originate behind the surfaces so as to
avoid singularities at $z=0,D$.

The second term $F_{S_1}$ on the RHS of Eq.~(\ref{eq3})
is the local contribution from the surface $S_1$ located
at $z=0$. We have written $\vec{r} = (\vec{\rho}, z)$, where
$\vec{\rho}$ denotes the $(d-1)$ coordinates parallel to the surface,
and $z$ denotes the coordinate perpendicular to the surface. In
$F_{S_1}$, $g$ and $\gamma$ are parameters which depend
on temperature and the exchange couplings in the bulk ($J$) and
at the surface ($J_s$) \cite{10,13}:
\begin{eqnarray}
\label{g}
g &=& \frac{(q - 2)J_s+J-k_BT}{k_B (T_c-T)} , \\
\label{gamma}
\gamma &=& \frac{J}{2 \xi_b k_B (T_c-T)} , \quad \xi_b =
\left[\frac{q}{2} \left( 1-\frac{T}{T_c} \right) \right]^{-1/2} .
\end{eqnarray}
Here, $q$ denotes the coordination number of a site, and
$\xi_b$ is the bulk correlation length.
[Our normalization of $F[\psi]$ in Eq.~(\ref{eq3})
implies that all lengths are measured in units of $\xi_b$.]
 Further, the dimensionless surface field
in $F_{S_1}$ is $h_1=-V(0)$. The one-sided derivative appears in
$F_{S_1}$ due to the absence of neighboring atoms for $z<0$.
Similarly, the third term $F_{S_2}$ is the contribution from the
surface $S_2$ located at $z=D$, with $h_2=-V(D)$. For simplicity,
we assume that $J_{s_1} = J_{s_2} = J_{s}$, so that the parameter
$g$ is the same for both $S_1$ and $S_2$.

The corresponding dynamical model is obtained as follows. In the
bulk, the order parameter evolves according to the Cahn-Hilliard-Cook
(CHC) equation for a conserved order parameter \cite{1,2,3,4,5,40}:
\begin{eqnarray} \label{eq4}
\frac{\partial} {\partial t} \psi(\vec{r}, t) &=& - \vec{\nabla}
\cdot \vec{J} (\vec{r}, t) \nonumber \\
&=& \vec{\nabla} \cdot [{\nabla} \mu
(\vec{r}, t) + \vec{\theta} (\vec{r}, t)] \nonumber \\
&=& \vec{\nabla} \cdot \left[\vec{\nabla} \left(\frac{\delta
F}{\delta \psi}\right) + \vec{\theta} (\vec{r}, t) \right] .
\end{eqnarray}
Here, $\vec{J}(\vec{r}, t)$ is the current, and $\mu (\vec{r}, t)$
is the local chemical potential difference
between A and B. Further, $\vec{\theta} (\vec{r}, t)$
is a random noise term, to be specified below. Using the free
energy functional from Eq.~(\ref{eq3}) in Eq.~(\ref{eq4}), we obtain
\begin{eqnarray} \label{eq5}
\frac{\partial} {\partial t} \psi (\vec{r}, t) = \vec{\nabla} \cdot
\left\{ \vec{\nabla} \left[- \psi + \psi^3 - \frac{1}{2} \nabla^2 \psi +
V (z) \right] + \vec{\theta} (\vec{r}, t) \right\}, \quad 0<z<D .
\end{eqnarray}
We assume that the noise $\vec{\theta}$ is a Gaussian white noise,
obeying the relations
\begin{eqnarray}
\label{eq6}
\langle \vec{\theta} (\vec{r}, t) \rangle &=& 0 \quad , \\
\label{eq7}
\langle \theta_i (\vec{r}\;', t') \theta_j (\vec{r}\;'', t'')
\rangle &=& 2 \epsilon \delta_{ij} \delta (\vec{r}\;' - \vec{r}\;
'') \delta (t'-t'') \quad .
\end{eqnarray}
The dimensionless noise amplitude is a function of the
temperature \cite{13}
\begin{equation} \label{eq8}
\epsilon=\frac{1}{3} \left(\frac{T_c}{T} - 1\right)^{-2}
\xi^{-d}_b \quad .
\end{equation}

Eqs.~(\ref{eq4})-(\ref{eq5}) model the fact that the order
parameter (the total concentration) is conserved \cite{40}.
However, it is important to note that the surface value of the
order parameter is not a conserved quantity. We assume a
nonconserved relaxational kinetics (referred to as {\it Model A}
\cite{40}) for this quantity at $S_1$:
\begin{eqnarray} \label{eq9}
\tau_0 \frac{\partial}{\partial t} \psi (\vec{\rho}, 0, t) 
&=& - \frac{\delta F}{\delta \psi (\vec{\rho},0,t)} \nonumber \\
&=& h_1 + g \psi (\vec{\rho}, 0, t)+\gamma \frac{\partial\psi}{\partial z}
\bigg|_{_{z=0}} \quad ,
\end{eqnarray}
where $\tau_0$ sets the time scale of the nonconserved kinetics.
Since the surface value of the order parameter relaxes much
faster than the time scales of phase separation, we subsequently
set $\tau_0=0$. Finally, we observe that there is no current of
material across $S_1$. This is implemented via a no-flux boundary
condition:
\begin{equation} \label{eq10}
J_z (\vec{\rho},0,t) = - \left\{\frac{\partial}{\partial z}
\left[ - \psi +\psi^3 - \frac{1}{2} \nabla^2 \psi + V(z) \right] +
\theta _z \right\}_{z=0} = 0 \quad .
\end{equation}
The boundary conditions at $z=D$ are implemented in a similar
manner. For the sake of completeness, we present them here:
\begin{eqnarray}
\label{eq11}
0 &=& h_2 + g \psi (\vec{\rho}, D, t) - \gamma \frac{\partial \psi}
{\partial z} \bigg|_{z=D} \quad , \\
\label{eq12}
0 &=& \left\{\frac{\partial} {\partial z} \left[-\psi + \psi^3
-\frac{1}{2} \nabla^2 \psi + V (z) \right] + \theta (z) \right\}_{z=D}
\quad .
\end{eqnarray}

Eqs.~(\ref{eq5})-(\ref{eq12}) constitute our model for phase
separation in a film \cite{39}. This model has been presented
in the context of a film with flat parallel surfaces.
However, the adaptation to an arbitrary geometry is obvious:
boundary conditions like Eqs.~(\ref{eq9})-(\ref{eq12}) are implemented
on all available surfaces. For example, Aichmayer et al. \cite{41}
have used the appropriate generalization of this model to study
SDSD in a cylindrical geometry.

We have undertaken a Langevin
simulation of the above model, including the noise term, in order
to study phase separation in both symmetric and antisymmetric
films. We implemented an Euler-discretized version of
Eqs.~(\ref{eq5})-(\ref{eq12}) on cubic lattices of
size $L^2 \times D$ with $L=256$ and $D=5,10$. The discretization
mesh sizes were $\Delta x=1$ and $\Delta t=0.02$. We should
stress that these mesh sizes are rather coarse and the resultant
numerical solution does not closely shadow the ``actual'' solution
of Eqs.~(\ref{eq5})-(\ref{eq12}). However, Oono and Puri \cite{42}
and Rogers et al. \cite{red88} have demonstrated that such discrete
{\it cell-dynamical system} models capture the physics of the
segregation process rather well. The boundary conditions in 
Eqs.~(\ref{eq9})-(\ref{eq12}) were implemented at
$z=0$ and $z=D$, respectively, while periodic boundary conditions
were applied in the $x$- and $y$-directions.

The nature of the surface potential $V(z)$ and the parameters
$g, \gamma, \epsilon$ determine the equilibrium phase
diagram of the film. For the semi-infinite case, the phase
diagram has been discussed in Refs.~\cite{pb92,10,13}. For the
films considered here, we have determined the boundary between
PW and CW phases both analytically
and numerically. For the sake of brevity, we do not discuss these
phase diagrams here, but point out that parameter values
are chosen to study ordering to both PW and CW states
in symmetric and antisymmetric films. The values of the
parameters in the boundary conditions were chosen as $g=-0.4$,
$\gamma=0.4$ \cite{13}. The potentials $V(z)$ which
we considered will be specified at appropriate places in the
subsequent discussion. The noise amplitude was fixed as
$\epsilon=0.327$, which corresponds to a quench with $T
\simeq 0.38~T_c$ from Eq.~(\ref{eq8}). The presence of thermal
fluctuations prevents the system from becoming stuck in metastable
configurations. However, we should stress that the ordering dynamics
is expected to be independent of noise in the asymptotic regime
\cite{po88,ab89}.

The initial conditions for our simulations consisted of a
homogeneous mixture of 50\% A and 50\% B, i.e.,
$\psi(\vec{r},t=0)=0 + \mbox{small-amplitude fluctuations}$.
This mimics the disordered high-temperature state for a
mixture with critical composition. We will characterize the
far-from-equilibrium dynamics of the quenched system via
evolution snapshots, laterally averaged order parameter profiles,
layer-wise correlation functions and length scales.
All statistical quantities were obtained as averages over
five independent runs.

\section{Kinetics of Phase Separation in a Symmetric Film}

Let us first consider the case of a symmetric film with a
power-law potential:
\begin{equation} \label{eq13}
V (z) =- V_0 \left[(z + 1)^{-3} + (D + 1-z)^{-3} \right] \quad .
\end{equation}
Recall that the exponent $n=3$ corresponds to non-retarded
van der Waals' interactions between a surface and a particle in
$d=3$. As the other parameters are fixed, an appropriate choice
of $V_0$ and $D$ results in either PW or CW states in equilibrium.
We have ascertained the PW-CW phase boundary by
studying the evolution of an initial condition which consists
of A-rich and B-rich domains separated by an A-B interface
along the $z$-direction. The onset of the CW phase is signaled
by the intrusion of a thin wetting layer [see Fig.~\ref{fig1}(b)]
between the B-rich domain and the surfaces. We will consider
the PW and CW cases separately.

\subsection{Partially Wet Surfaces}

\begin{figure}[htb]
\includegraphics[width=110mm]{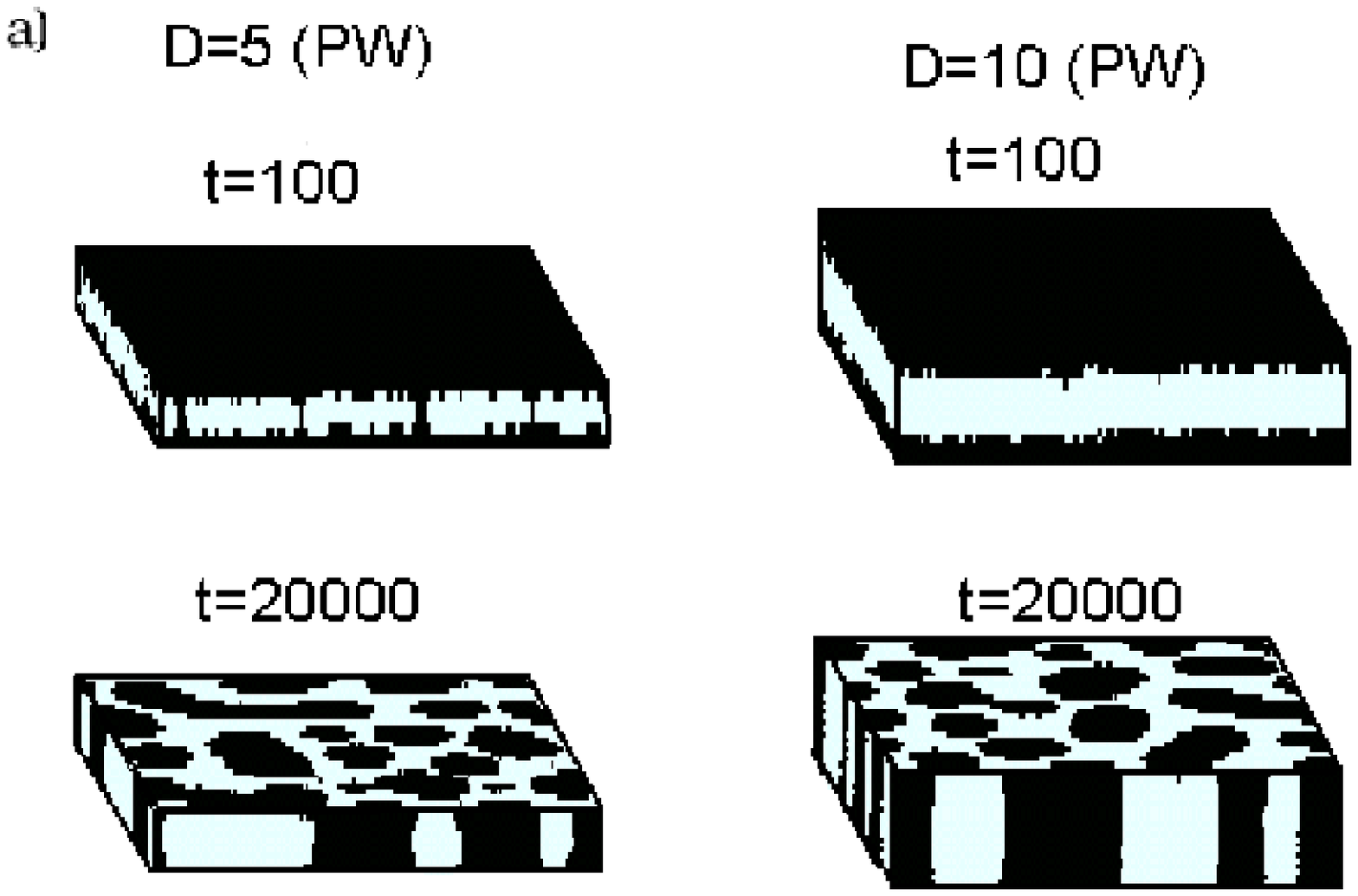}
\vskip0.4cm
\includegraphics[width=110mm]{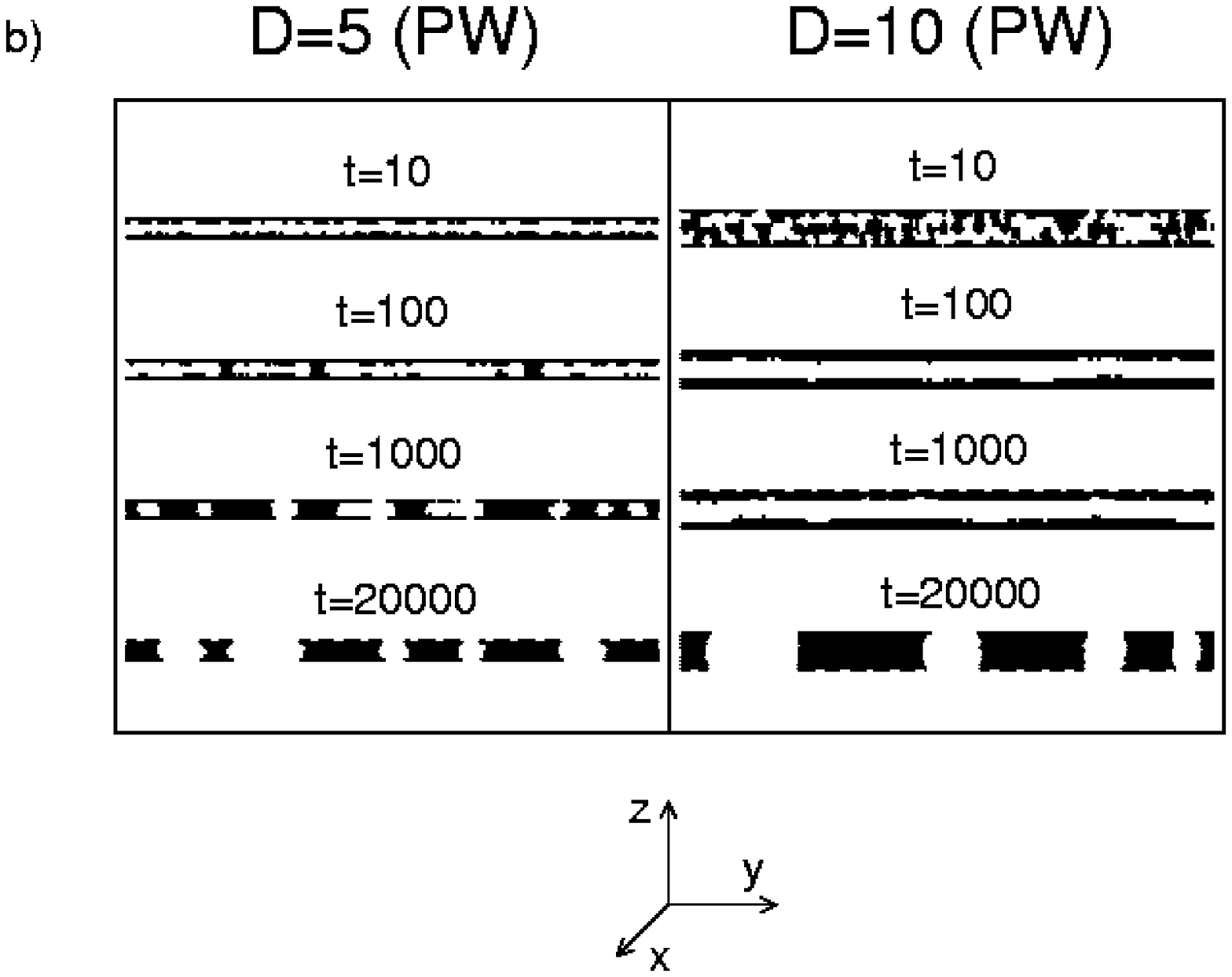}
\caption{\label{fig2} (a) Evolution snapshots at times $t=100$ and
$t=20000$ for a critical binary mixture in a symmetric film with
PW morphology. The system
size was $L^2 \times D$ with $L=256$ and $D=5$ (frames on left)
and $D=10$ (frames on right). The A-rich regions are
colored black, and the B-rich regions are colored
white (light blue online). (b) Perpendicular cross-sections of the
snapshots in (a) at $y=L/2$ in the $(xz)$-plane. The A-rich
regions are marked black, and the B-rich regions are unmarked.}
\end{figure}

In Fig.~\ref{fig2}, we show evolution snapshots [part (a)] and
($xz)$-cross-sections [part (b)] for films with $D=5$ (frames on
left) and $D=10$ (frames on right). The potentials were chosen
with $V_0=0.325$ for $D=5$ and $V_0=0.11$ for $D=10$, which
correspond to the PW state in equilibrium. (The critical value
of $V_0/\sigma$ for the PW$\rightarrow$CW crossover diminishes
with increase in $D$, and $V_0/\sigma \rightarrow 1$ as
$D \rightarrow \infty$.) Note that a metastable layered
structure forms at early times, since the kinetics
of surface enrichment \cite{43} is much faster than the time scale
of phase separation. On longer time scales, spinodal fluctuations
break this layered structure and the system forms domains which
coarsen in directions parallel to the surface. We stress that
the layered state can be very long-lived, and may be misinterpreted
as evidence for the formation of wetting layers in experiments.

\begin{figure}
\includegraphics[width=120mm]{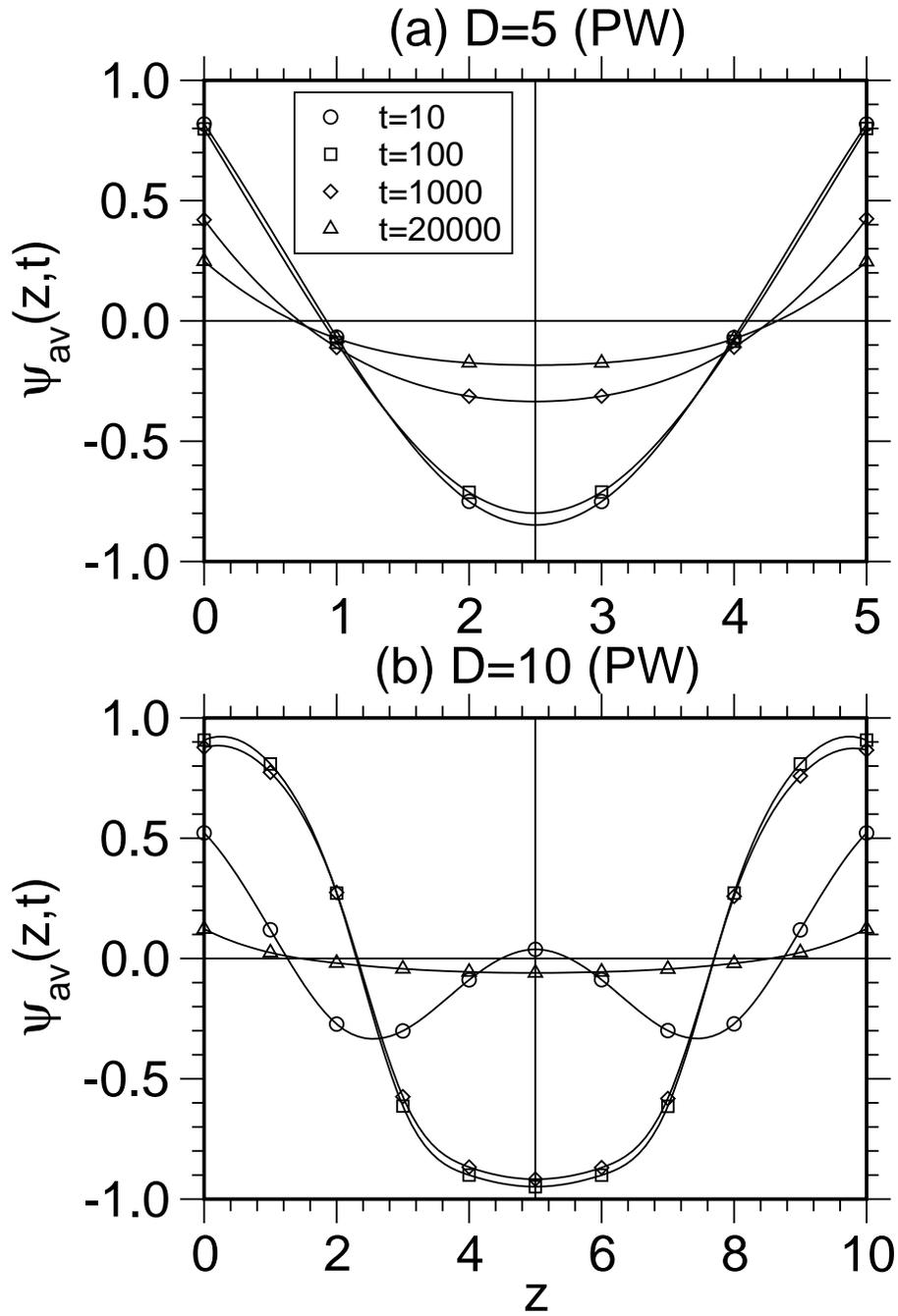}
\caption{\label{fig3} Laterally averaged profiles for the
evolution depicted in Fig.~\ref{fig2} at the dimensionless times
$t=10,100,1000,20000$, for (a) $D=5$, and (b) $D=10$. The symbols
denote the same times in both figures.}
\end{figure}

Many experimental probes (such as depth-profiling techniques) do
not have any lateral resolution, and yield only laterally averaged
order parameter profiles $\psi_{\rm av} (z,t)$ vs. $z$ \cite{9,12}.
In our simulations, laterally averaged profiles are obtained by
averaging $\psi (x,y,z,t)$ along the $x$- and $y$-directions, and then
further averaging over five independent runs. The depth profiles
corresponding to the evolution in Fig.~\ref{fig2} are shown in
Fig.~\ref{fig3}. For bulk spinodal
decomposition, the wave-vectors are randomly oriented and the
averaging procedure yields $\psi_{\rm av} (z,t)\simeq 0$.
For SDSD, however, the laterally averaged profiles are systematic
at the surfaces since the boundary conditions result in
spinodal waves with wave-vectors perpendicular to
the surfaces. Let us focus on the
case with $D=10$ in Fig.~\ref{fig3}(b). The profile at time $t=10$
shows the formation of two symmetric SDSD waves, which propagate
towards the center of the film. The $t=100$ profile shows the
metastable layered state that has originated from
these waves. This structure is also present at $t=1000$, and may
be misinterpreted as a CW equilibrium state which occurs for
temperatures between the critical temperature of
the thin film and the critical
temperature of the bulk system \cite{31,32}. Finally, the
spinodal fluctuations break this structure and the averaged
profile at $t=20000$ is almost flat. Since a weak surface
field amplitude $V_0=0.11$ was chosen in this case, there is only a
slightly A-rich region [$\psi_{\rm av}(z,t) > 0$] near the walls and,
correspondingly, only a slightly A-poor region [$\psi_{\rm av}(z,t) <
0$] at the center.

\begin{figure}
\includegraphics[width=150mm]{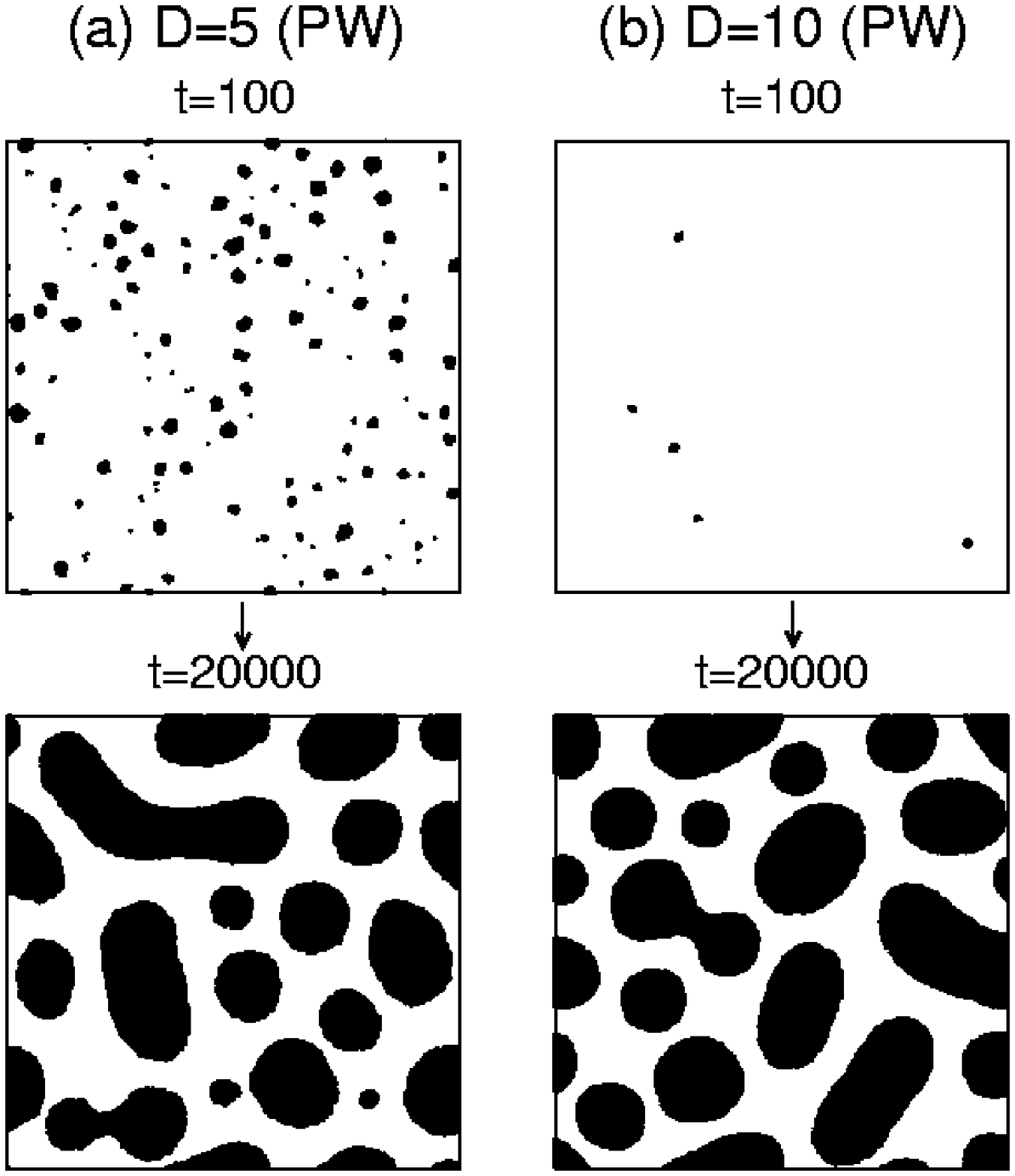}
\caption{\label{fig4} Cross-sections of the evolution snapshots
in Fig.~\ref{fig2}(a). The cross-sections are taken parallel to the
surfaces at (a) $z=2$ for $D=5$, and (b) $z=5$ for $D=10$.}
\end{figure}

It is also interesting to study $(xy)$-cross-sections of the
evolution snapshots in Fig.~\ref{fig2}. In Fig.~\ref{fig4}, we
show the relevant cross-sections at $z=2$ for $D=5$ and
$z=5$ for $D=10$. For early times ($t=100$) the central region is
strongly depleted in A due to the formation of the layered
structure. The resultant morphology corresponds
to an off-critical composition with droplets of A in
a matrix of B. At later times, $t=20000$, the central region has
almost equal amounts of A and B again. However, there is still
a small depletion in A (see Fig.~\ref{fig3}), and hence
the growth morphology still contains droplets of A.

Let us next focus on the layer-wise correlation function, which is
defined as \cite{39}
\begin{equation} \label{eq14}
C_{_{\parallel}} (\vec{\rho}, z,t) =L^{-2} \int d \vec{\sigma}~
[\langle \psi(\vec{\sigma}, z,t) \psi (\vec{\sigma} + \vec{\rho},
z,t) \rangle - \langle \psi(\vec{\sigma}, z,t) \rangle \langle
\psi(\vec{\sigma} + \vec{\rho}, z,t) \rangle ] \quad ,
\end{equation}
where the angular brackets denote an averaging over independent
runs. Since the system is isotropic in the ($xy$)-plane,
$C_{_{\parallel}}$ does not depend on the direction of
$\vec{\rho}$. We can define the $z$-dependent lateral length scale
$L_{{_\parallel}} (z,t) \equiv L(z,t)$ from the half-decay of
$C_{{_\parallel}}(\rho, z, t)$ \cite{39}:
\begin{equation} \label{eq15}
C_{_{\parallel}} (\rho=L, z,t) = \frac{1}{2} C_{_{\parallel}} (0,z,t).
\end{equation}

\begin{figure}
\includegraphics[width=150mm]{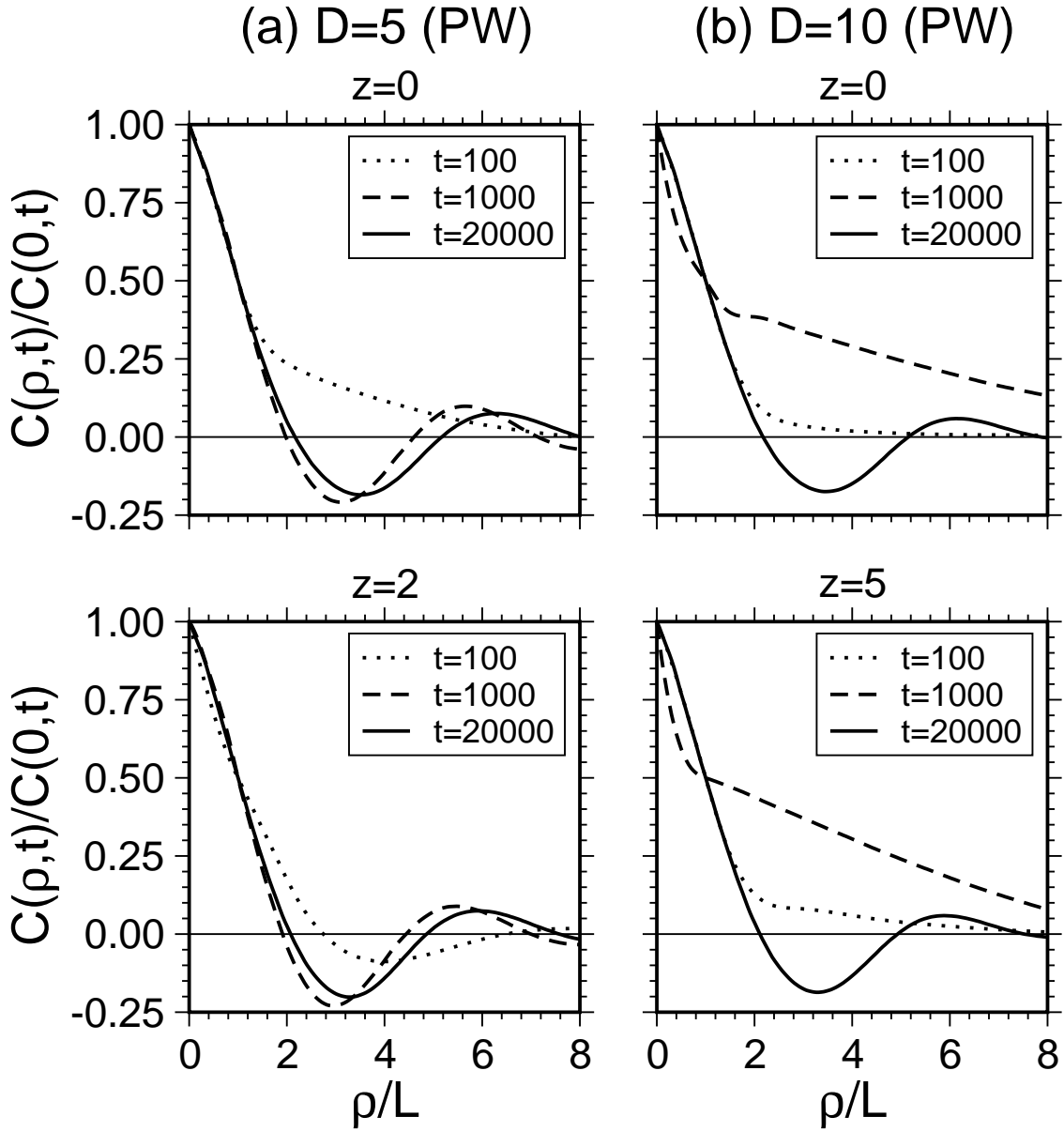}
\caption{\label{fig5}Scaling plot of layer-wise
correlation functions, defined in Eq.~(\ref{eq14}), for the
evolution depicted in Fig.~\ref{fig2}. We plot data for
$C(\rho, t)/C(0,t)$ vs. $\rho/L$ for three different times,
$t=100,1000,20000$. We present data for (a) $D=5$ at $z=0$
(wall) and $z=2$ (center); and (b) $D=10$ at $z=0$ (wall) and
$z=5$ (center). The layer-wise length-scale $L(z,t)$ is defined
as the distance over which $C(\rho,t)$ has decayed to 1/2 its
maximum value (at $\rho = 0$).}
\end{figure}

For convenience, we denote $C_{_{\parallel}} (\rho, z, t)$
as $C(\rho,t)$ in the following discussion.
In Figs.~\ref{fig5}(a),(b) we plot
the scaled correlation functions $C(\rho, t)/C(0,t)$ vs. $\rho/L$
for $D=5$ and $D=10$, respectively. In bulk systems, the correlation
function exhibits {\it dynamical scaling}, viz., $C(\vec{r},t) =
g(r/L)$, where $g(x)$ is independent of time. This property
indicates that the evolution morphology is statistically self-similar
in time, and only the scale of the morphology changes. In this case,
there is no dynamical scaling as the correlation functions correspond
to qualitatively different morphologies (see Fig.~\ref{fig2}).
Thus, for $D=10$ and $t=1000$, the layered structure has not
yet broken up, while at $t=20000$ lateral phase separation has
occurred. Of course, dynamical scaling is recovered subsequent
to the formation of well-formed laterally segregated domains
($t \geq 10000$ for both $D=5$ and 10).

\begin{figure}
\includegraphics[width=120mm]{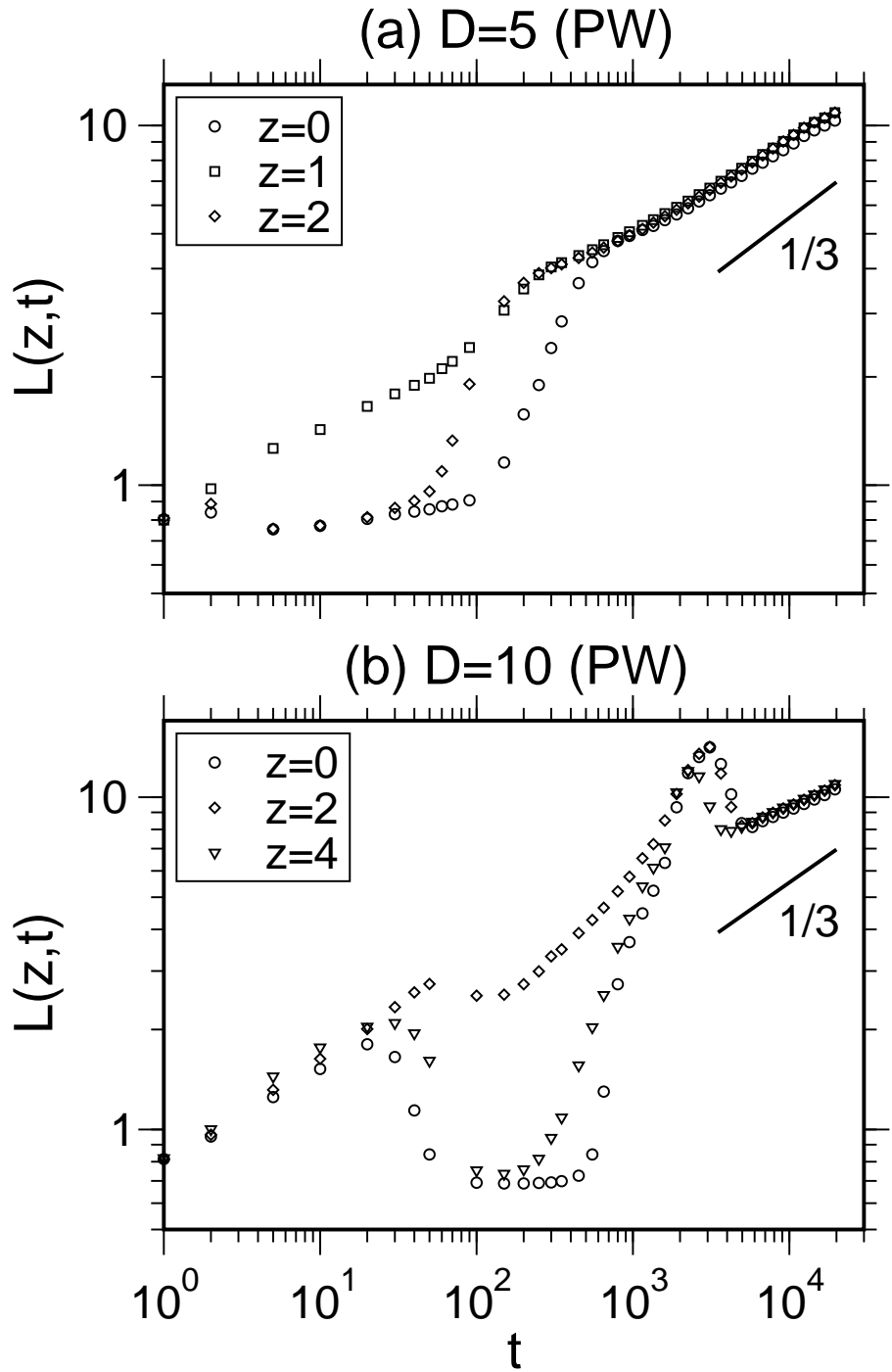}
\caption{\label{fig6} Time-dependence of the layer-wise
length scale for the evolution depicted in Fig.~\ref{fig2}. We
plot $L(z,t)$ vs. $t$ on a log-log scale for various values of $z$
and (a) $D=5$, and (b) $D=10$. The lines of slope
1/3 denote the Lifshitz-Slyozov (LS) growth law, $L(t) \propto
t^{1/3}$, which characterizes diffusion-driven phase separation in
the bulk.}
\end{figure}

Finally, we examine the time-dependence of the lateral domain size
$L(z,t)$ in Fig.~\ref{fig6}. While the asymptotic growth is
consistent with the expected Lifshitz-Slyozov (LS) growth law
$L(t) \sim t^{1/3}$ \cite{1,2,3,4,5}, which describes bulk domain
growth, the early-time dynamics is complicated. For the
$D=10$ case in Fig.~\ref{fig6}(b), the early-time data corresponds
to the growth of bulk-like domains before the layered structure has
formed. The spinodal fluctuations originate in the central region
($z=5$) where the surface field is not felt, and propagate to the
surface ($z=0$). The break-up of the layered structure is
characterized by the non-monotonic behavior of $L(z,t)$ vs. $t$.
As the fluctuations originate near the film center, the data set for
$z=0$ is the last to become consistent with LS behavior.

\subsection{Completely Wet Surfaces}

\begin{figure}
\includegraphics[width=110mm]{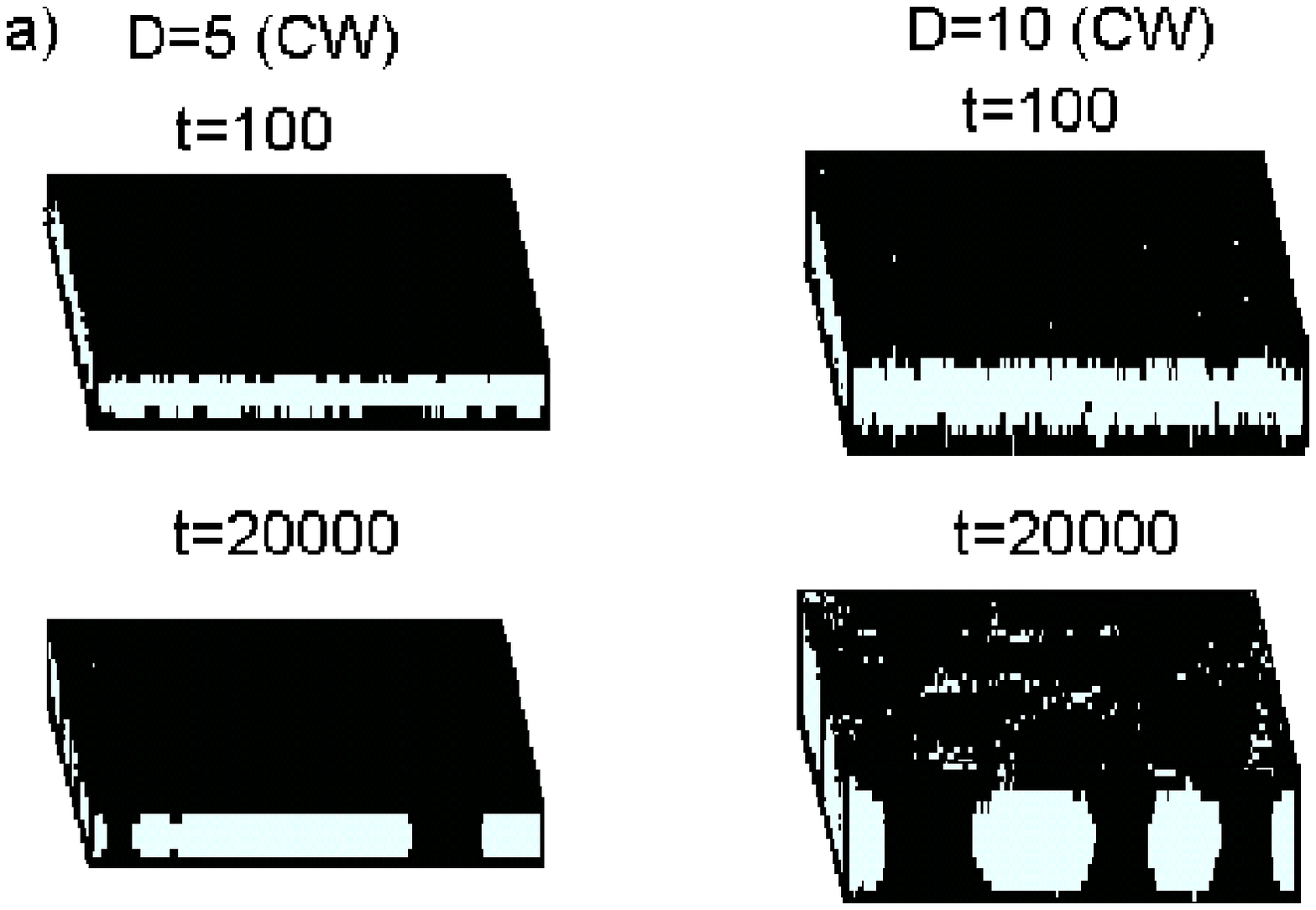}
\includegraphics[width=110mm]{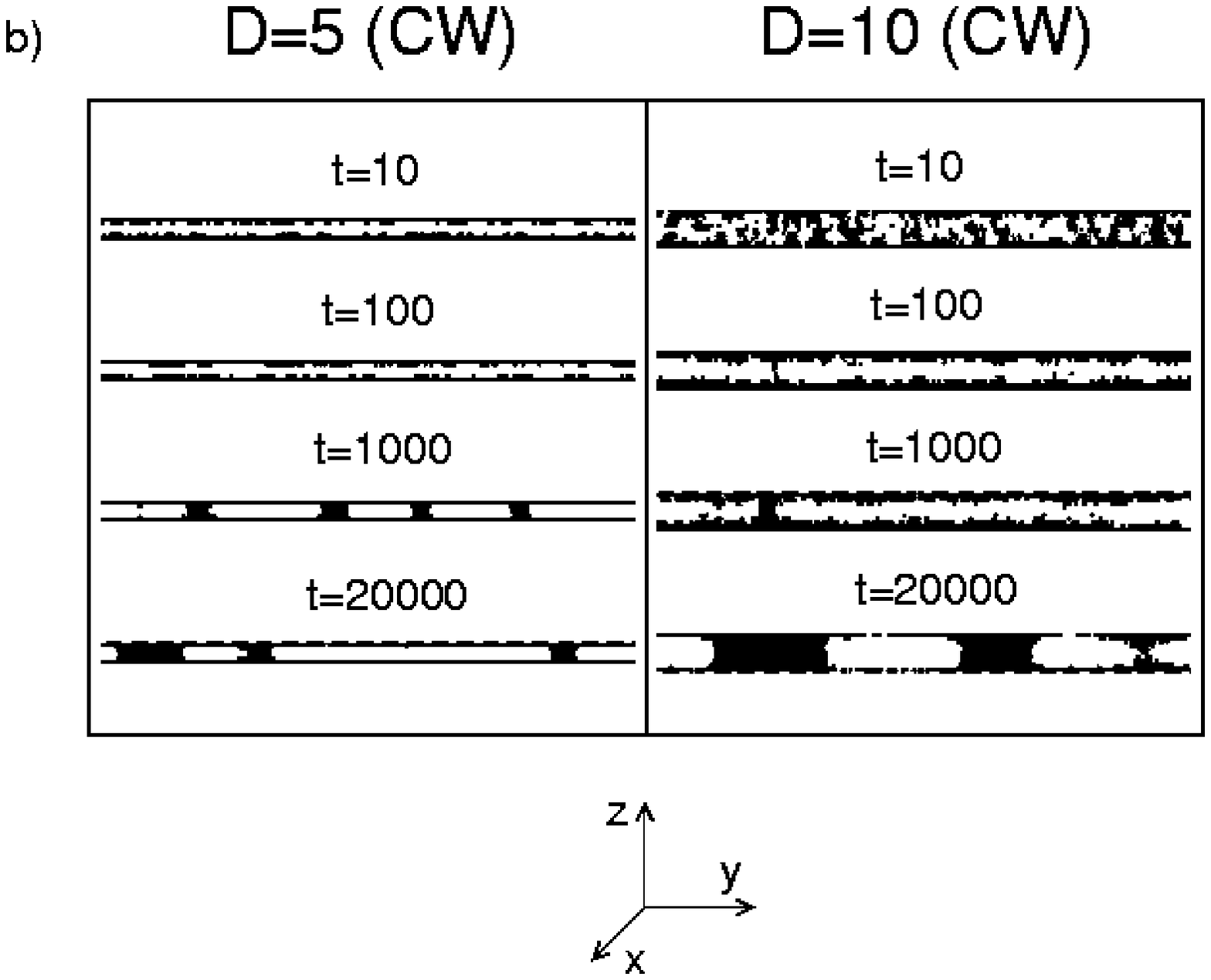}
\caption{\label{fig7}
Analogous to Fig.~\ref{fig2}, but for a symmetric film with
a CW morphology. The parameter values are just above the PW-CW
boundary for the $D=10$ case. Therefore, the surface does not
have a clean coating of the A-rich phase.}
\end{figure}

\begin{figure}
\includegraphics[width=120mm]{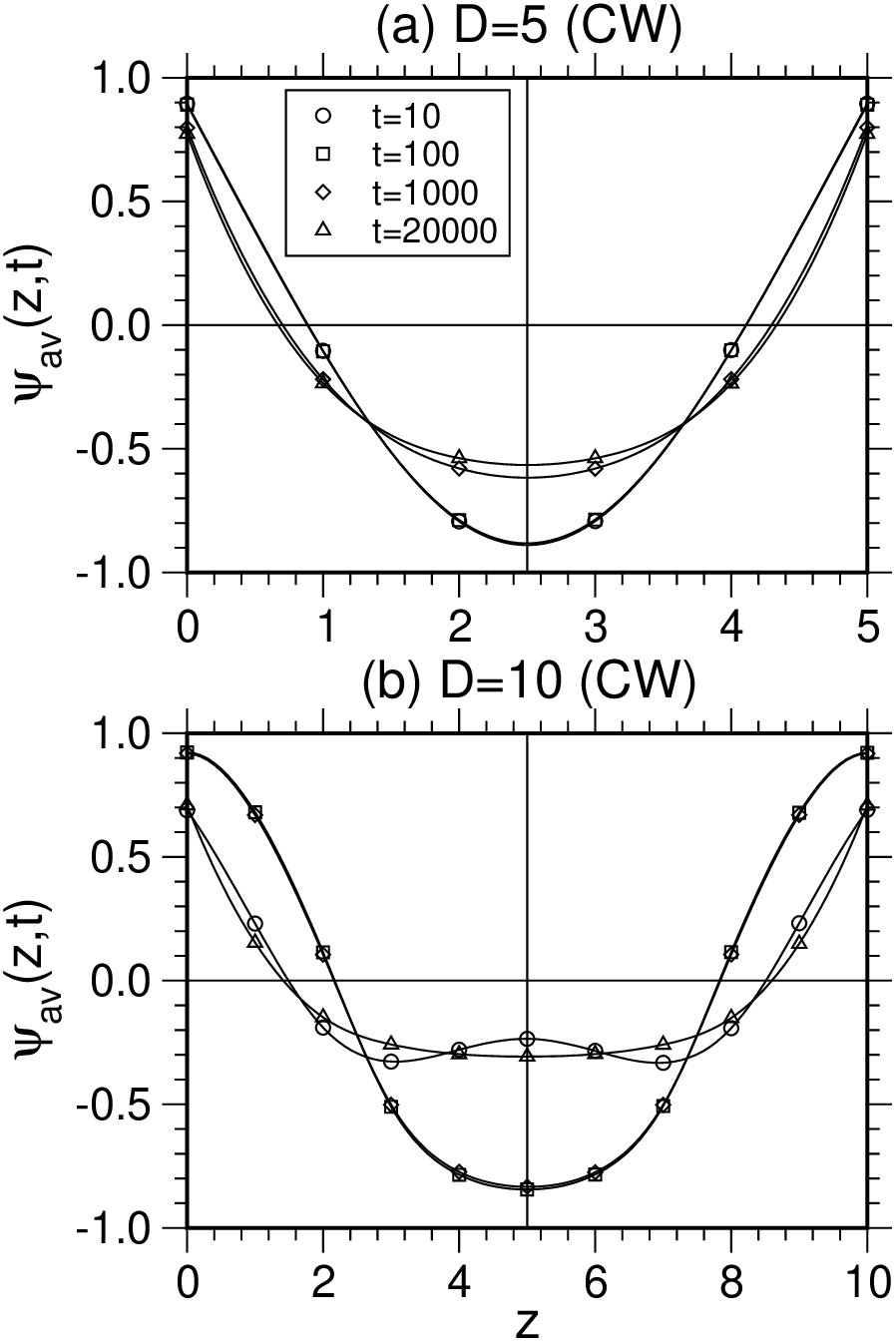}
\caption{\label{fig8} Laterally averaged profiles
for the evolution depicted in Fig.~\ref{fig7} at the dimensionless
times $t=10, 100,1000,20000$, for (a) $D=5$, and (b) $D=10$.}
\end{figure}

Next, let us consider the case where the surfaces have a
CW morphology in equilibrium. In Fig.~\ref{fig7}, we show
evolution snapshots and $(xz)$-cross-sections for the CW case.
The corresponding potential strengths were $V_0=0.45$ for $D=5$,
and $V_0=0.275$ for $D=10$. Again, the system forms a metastable
layered structure at early times, which is broken up by
spinodal fluctuations at later times. (For very strong surface
fields, the layered structure actually corresponds to an
equilibrium state and the corresponding pattern dynamics
is uninteresting.) However, the difference
in this case is that the B-rich regions are encapsulated by A
[see Fig.~\ref{fig1}(b)], unlike the situation shown in
Fig.~\ref{fig2}. The asymptotic dynamics is then characterized by
the coarsening of these encapsulated pancakes. The laterally
averaged profiles (Fig.~\ref{fig8}) again show that the initial
layered structure is rather pronounced (compare with
Fig.~\ref{fig3}). The depth profiles become softer at later times,
but due to the pancakes being encapsulated by A, there remains
a strong surface enrichment in A. If one looks at
cross-sections taken parallel to the surfaces, analogous to
Fig.~\ref{fig4}, one finds a qualitatively similar behavior. Of
course, the volume fraction of A in the central region is now
smaller due to the higher degree of surface enrichment.

\begin{figure}
\includegraphics[width=110mm]{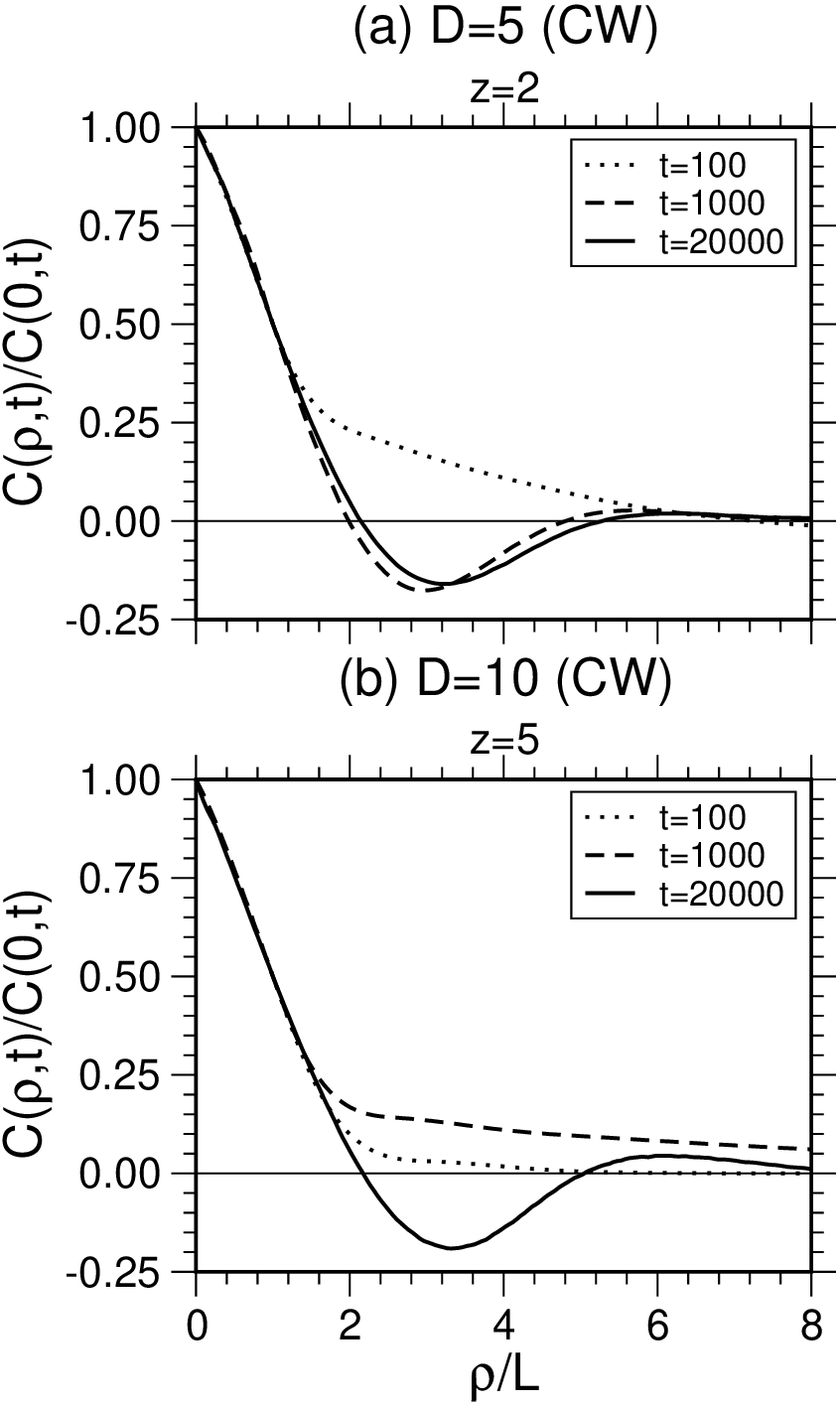}
\caption{\label{fig9} Scaling plot of layer-wise correlation
functions for the evolution depicted in Fig.~\ref{fig7}. We plot
$C(\rho,t)/C(0,t)$ vs. $\rho/L$ for $t=100,1000,20000$.
We present data for (a) $D=5$ at $z=2$ (center); and (b) $D=10$ at
$z=5$ (center).}
\end{figure}

Figure~\ref{fig9} is a scaling plot of $C(\rho,t)/C(0,t)$ vs.
$\rho/L$, and is analogous to Fig.~\ref{fig5}. We do not show data
for the $z=0$ case here as the surface is always A-rich and does
not exhibit interesting pattern dynamics. The behavior in the film
center is qualitatively similar to the PW case, i.e., there is no
dynamical scaling for the time range shown. This can be
understood in the context of the evolution dynamics shown in
Fig.~\ref{fig7}(b). For the $D=10$ case, the morphologies exhibit
a crossover behavior from the layered state to the (asymptotic)
pancake state for $t=100,1000,20000$. For $t \geq 20000$, we expect
to recover dynamical scaling. For the $D=5$ case, the system is
almost in its asymptotic state by $t=1000$. Hence, the correlation
functions for $t=1000$ and $t=20000$ show approximate scaling.

In Fig.~\ref{fig10}, we study the time-dependence of the parallel
length scale $L(z,t)$. The non-monotonic behavior again reflects
the formation and break-up of a long-lived metastable layered
structure. Notice that this state is far from equilibrium
despite the fact that we do expect a CW morphology for these
parameter values. The break-up of the layered structure gives rise
to the growth of laterally segregated domains. Although we have
followed the growth of $L(z,t)$ over several decades in time, the
expected asymptotic regime of LS growth is not observed over
simulation time-scales.

\section{Kinetics of Phase Separation in an Antisymmetric Film}

We next consider the case of an antisymmetric film. The
corresponding power-law potential is
\begin{equation} \label{eq16}
V(z) = - V_0 \left[(z+1)^{-3} - (D + 1 -z)^{-3} \right] ,
\end{equation}
so that $V(D-z) = - V(z)$. In Figs.~\ref{fig1}(c)-(d), we have
schematically shown the PW and CW states which arise for an
antisymmetric film. In this case also, we have
obtained the PW-CW phase boundary as a function
of $V_0$ and $D$. As before, we will consider and compare
both PW and CW cases.

\subsection{Partially Wet Surfaces}

In Fig.~\ref{fig11}, we show evolution snapshots [part (a)] and
($xz)$-cross-sections [part (b)] for films with $D=5$ (frames on
left) and $D=10$ (frames on right). The potential strengths were
$V_0=0.055$ for $D=5$ and $V_0=0.041$ for $D=10$,
which correspond to a PW case in equilibrium. In
the $D=5$ case, we observe the formation of a layered state which
breaks up into a coarsening domain structure. A similar evolution
occurs in the $D=10$ case, though the layered state (at $t=1000$)
is not so clean for these weak surface fields. However, it shows
up more clearly in the laterally averaged profiles we present
next. The domain cross-sections
are trapezoidal with different contact angles at the lower
surface (which prefers A) and the upper surface (which prefers B).

The laterally averaged profiles (Fig.~\ref{fig12}) confirm that
a layered state, with a single interface, appears as a transient
before the lateral domain growth sets in. For the $D=10$ case,
we see the interaction of two opposite SDSD waves (at $t=10,100$),
resulting in the formation of the layered state (at $t=1000$).
In Fig.~\ref{fig13},
we examine the morphology in planes parallel to the surfaces. For the
$D=5$ case, we focus on a cross-section at $z=2$. The $t=100$
morphology corresponds to the layered state [see Fig.~\ref{fig12}(a)]
and consists of droplets of B in a matrix of A. The $t=20000$
morphology corresponds to the laterally-segregated state and
consists of bicontinuous domains. For the $D=10$ case, we consider a
cross-section at $z=5$, which is precisely the film center.
Since the film always has a near-critical composition at the center
[see Fig.~\ref{fig12}(b)], the segregation morphology is bicontinuous
for both $t=100$ and $t=20000$ in this case.

We have also examined the layer-wise
correlation functions $C_{_{\parallel}} (\vec{\rho},z,t) \equiv
C(\rho,t)$ (Fig.~\ref{fig14}), and the lengths $L(z,t)$ which
one can extract from them (Fig.~\ref{fig15}). Again,
one finds marked deviations from scaling for $C(\rho, t)$, as
expected due to the transient formation of layered structures.
Of course, scaling is recovered in the asymptotic regime,
which is characterized by the coarsening of trapezoidal domains.
In the $D=5$ case, the length-scale data shows that the asymptotic
behavior is consistent with the LS law, $L(t) \sim t^{1/3}$.
A similar behavior is seen for the film with $D=10$, though the
non-monotonic behavior is less pronounced in this case. This
emphasizes that one has to be careful with the interpretation of
growth phenomena in confined geometries, and rather complete
information on the structural evolution of a system is mandatory
for establishing a clear picture.

\subsection{Completely Wet Surfaces}

Our last set of numerical results corresponds to the CW state for
an antisymmetric film: in this case, a layered state with a
single interface is the equilibrium state [see Fig.~\ref{fig1}(d)],
and no lateral segregation
should occur! In Fig.~\ref{fig16}, we show evolution snapshots and
$(xz)$-cross-sections for $D=5$ with $V_0 = 0.25$ (frames on left)
and $D=10$ with $V_0 = 0.2$ (frames on right). There is seen to be
some lateral inhomogeneity in the early stages. Starting from a random
initial state, small A-rich and B-rich domains are formed first.
It takes time for the interfaces
between these small domains to annihilate by diffusion and
coalescence, until a single domain wall parallel to the surface is
left. The corresponding laterally averaged profiles are shown in
Fig.~\ref{fig17}. In particular, we draw the reader's attention to
Fig.~\ref{fig17}(b), which shows the formation and collision of two
opposite SDSD waves originating from $S_1$ and $S_2$ -- see
profiles for $t=10$ and $t=100$. Figure~\ref{fig18} shows the
evolution pictures at the film center ($z=5$) for the $D=10$ case.
(The corresponding pictures at $z=2$ or $z=3$ for the $D=5$
case merely show a uniform state.) The length scale at $z=5$ for
$D=10$ (Fig.~\ref{fig19}) grows uniformly for $t \geq 1000$, and
the time-dependence is consistent with the LS law.

\section{Summary and Discussion}

Let us conclude this paper with a summary and discussion of the
results. We have studied the diffusion-driven phase separation of an
AB mixture confined in a film. The film has two parallel surfaces
$S_1$ and $S_2$, which are separated by a distance $D \sim
\mbox{O}(10 \xi_b)$, where $\xi_b$ is the bulk correlation length.
We have considered (a) {\it symmetric
films}, where $S_1$ and $S_2$ have an identical attraction
for the A-component; and (b) {\it antisymmetric films}, where $S_1$ and
$S_2$ attract (with equal strength) the A-component and B-component,
respectively. Both cases are of considerable experimental relevance.

The equilibrium segregated state can be either partially wet (PW)
or completely wet (CW), depending on the nature of the surface
potentials. Further types of mixed-phase states in thin films occur
only for restricted ranges of parameters \cite{34,35,36},
and are not considered in the present paper. We have
clarified the typical growth scenario in both symmetric and
antisymmetric films. In both PW and CW cases, the surfaces give
rise to surface-directed spinodal decomposition (SDSD) waves,
which propagate towards the film center. The interaction of these
SDSD waves leads to the formation of a layered state. This
state is metastable for the PW case, and is broken up by
spinodal fluctuations. (However, the metastable state may have a
very long lifetime, and could be of considerable experimental
significance.) The subsequent evolution of the mixture is
characterized by the lateral coarsening of pancake-like domains.
(For antisymmetric films, these domains are trapezoidal because of
the different contact angles at $S_1$ and $S_2$.)
In the later stages, we expect that this coarsening is governed by
the Lifshitz-Slyozov (LS) growth law $L(t) \sim t^{1/3}$, but
one often encounters slow transients and even a non-monotonic
variation of $L(t)$ with time, before the LS regime sets in.
For the CW case in symmetric films, the initial layered state is
again metastable for moderate surface fields,
and breaks up into encapsulated pancakes
which coarsen in the lateral direction. For the CW case in
antisymmetric films, the equilibrium state is a layered state with
a single interface parallel to the surfaces. The system can relax
to this state rather rapidly.

At this stage, one may ask what happens if hydrodynamic effects
are incorporated into the above discussion. It is
well-known \cite{1,2,3,4,5} that bulk fluid mixtures exhibit
more complicated segregation kinetics than solid mixtures. In the
initial stages, growth is diffusive and is governed by the LS
growth law. However, at later times, hydrodynamic effects become
relevant and domain growth is facilitated by advective transport
along interfaces. The corresponding growth laws are $L(t) \sim t$
in the {\it viscous hydrodynamic} regime \cite{es79}, and $L(t)
\sim t^{2/3}$ in the {\it inertial hydrodynamic} regime \cite{hf85}.
The effects of surfaces on phase-separating binary fluids can be
studied at various levels of description. At the coarse-grained level,
an appropriate model is {\it Model H} at a surface \cite{40,44}.
This consists of the coupled dynamics of an order-parameter field
and a velocity field, with appropriate boundary conditions at the
surfaces. Alternatively, one can study mesoscale models consisting,
e.g., of evolution equations for the configuration probability
distribution \cite{bpl01}. Finally, at the microscopic level, one
can undertake molecular dynamics (MD) simulations of binary fluids
\cite{45} in a confined geometry. We have undertaken such MD
simulations, and will report the results in a forthcoming
publication.

More generally, we emphasize that there are many intriguing
aspects of phase separation in confined geometries, and a
number of possible directions for further investigation.
For example, the present study focused on mixtures with critical
composition. In a semi-infinite geometry, Puri and
Binder \cite{pb01} have demonstrated that novel features arise
when off-critical compositions are considered, e.g., the
wetting layer grows faster when the wetting component is a
minority phase rather than a majority phase. It would also
be interesting to study the phase separation of off-critical
mixtures in the present context of confined geometries.
Another interesting complication arises if there is frozen-in
disorder at the surfaces, which affects the wetting behavior.
Finally, it is also relevant to study SDSD in more complex
confined geometries than those studied here, e.g., wedges, patterned
surfaces, etc. Though the modeling of
these problems is straightforward, we expect that they will
give rise to novel physical phenomena. There remain many issues to
be addressed in this area, and we urge experimentalists to undertake
fresh experiments in these directions.

\vskip0.5cm
\noindent{\bf Acknowledgments}: SKD received financial
support from the Deutsche Forschungsgemeinschaft (DFG) via grant
No Bi314/18-2. JH acknowledges support from the Emmy
Noether Program at the DFG. SP is grateful to K. Binder for
inviting him to Mainz, where this work was initiated. Furthermore
support from the DFG via Sonderforschungsbereich 625/A3 is acknowledged.

\newpage
\begin{center}
{\bf Figures and Figure Captions}
\end{center}

\begin{figure}
\includegraphics[width=120mm]{fig10.eps}
\caption{\label{fig10} Time-dependence of the layer-wise
length scale for the evolution depicted in Fig.~\ref{fig7}. We
plot $L(z,t)$ vs. $t$ on a log-log scale for various values of $z$
and (a) $D=5$, and (b) $D=10$.}
\end{figure}

\begin{figure}
\includegraphics[width=110mm]{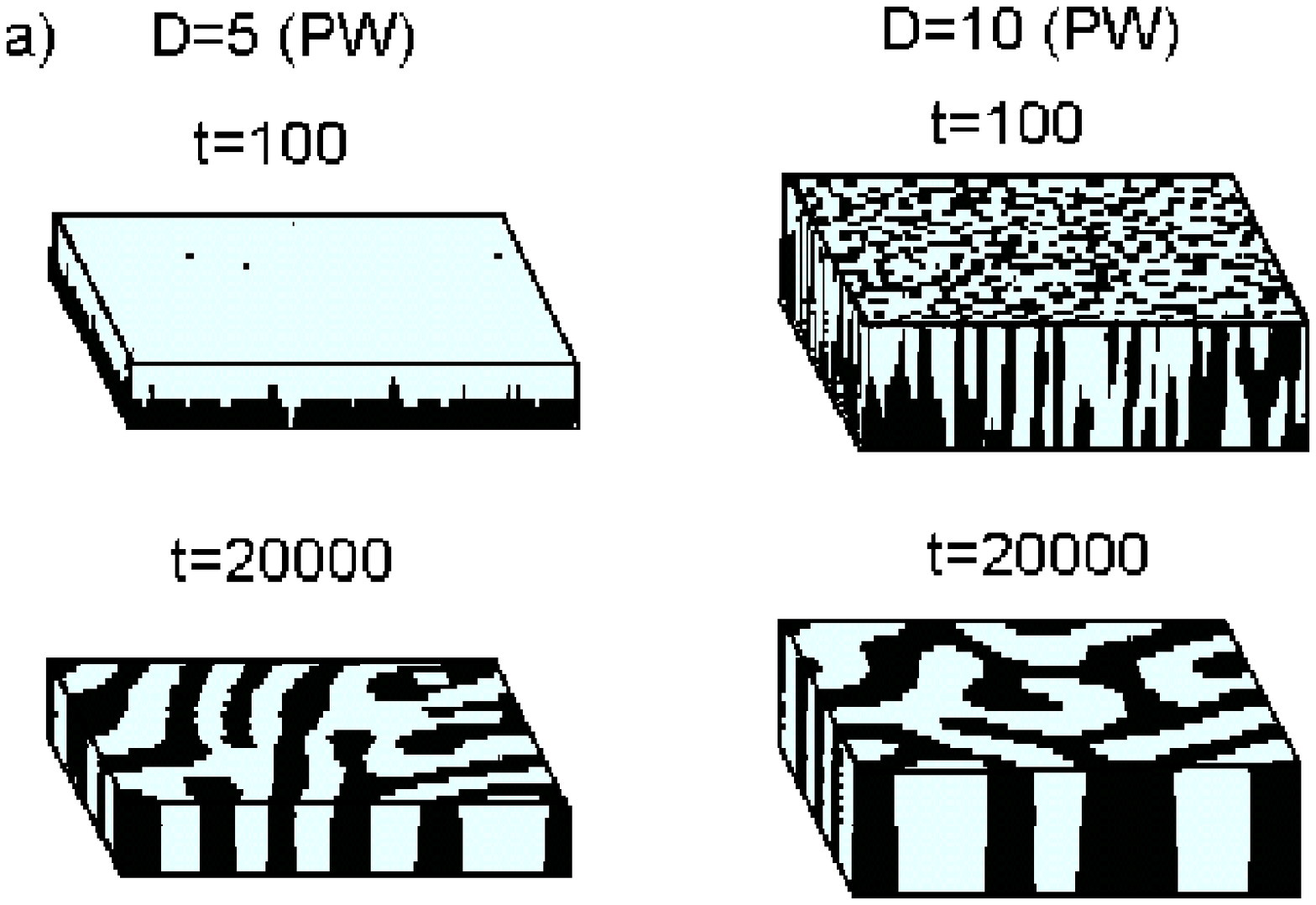}
\includegraphics[width=110mm]{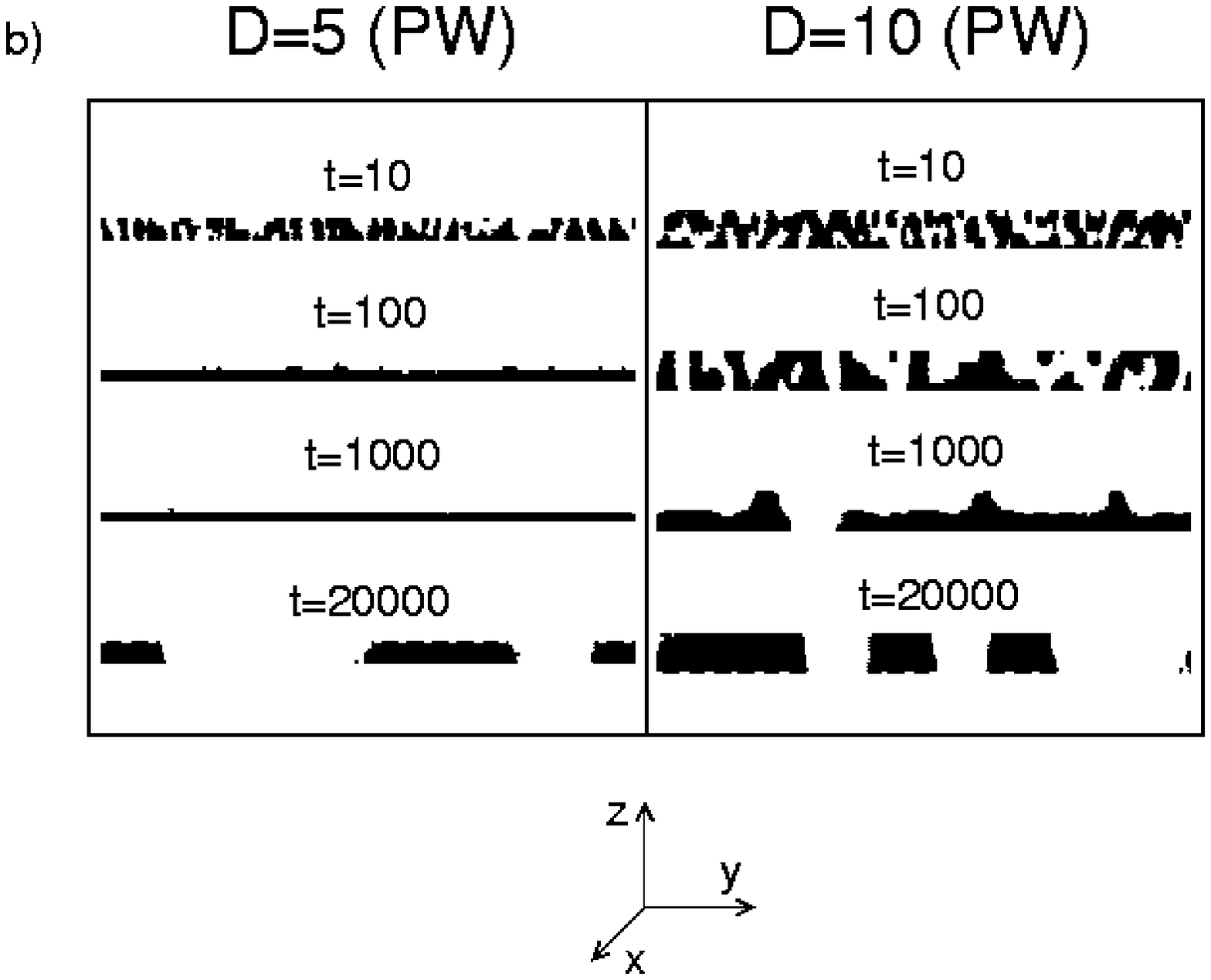}
\caption{\label{fig11}
Analogous to Fig.~\ref{fig2}, but for an antisymmetric film with
a PW morphology.}
\end{figure}

\begin{figure}
\includegraphics[width=120mm]{fig12.eps}
\caption{\label{fig12}Laterally averaged profiles for
the evolution depicted in Fig.~\ref{fig11} at the dimensionless
times $t=10,100,1000,20000$, for (a) $D=5$, and (b) $D=10$.}
\end{figure}

\begin{figure}
\includegraphics[width=150mm]{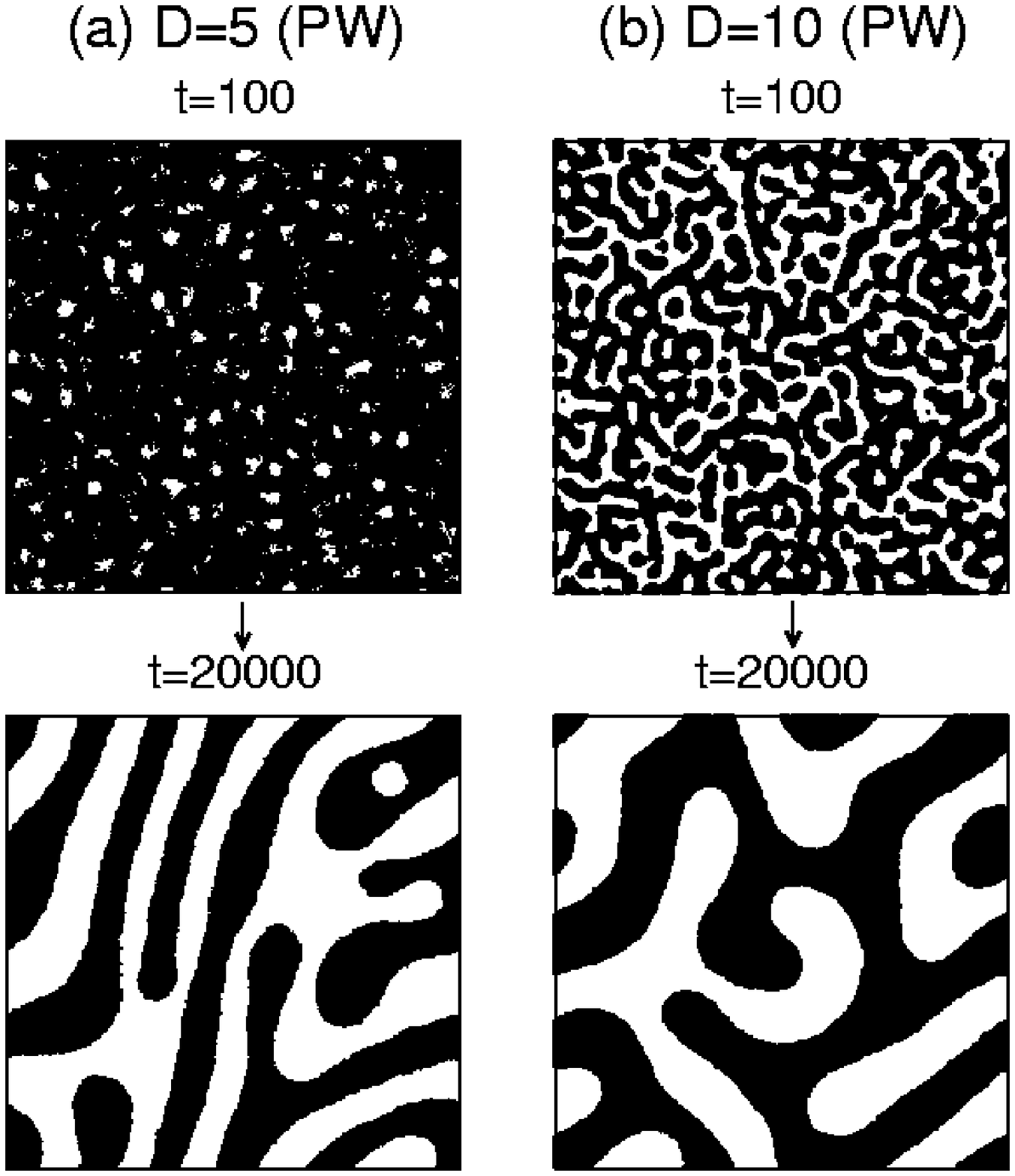}
\caption{\label{fig13} Cross-sections of the evolution snapshots
in Fig.~\ref{fig11}(a). The cross-sections are taken parallel to the
surfaces at (a) $z=2$ for $D=5$, and (b) $z=5$ for $D=10$.}
\end{figure}

\begin{figure}
\includegraphics[width=150mm]{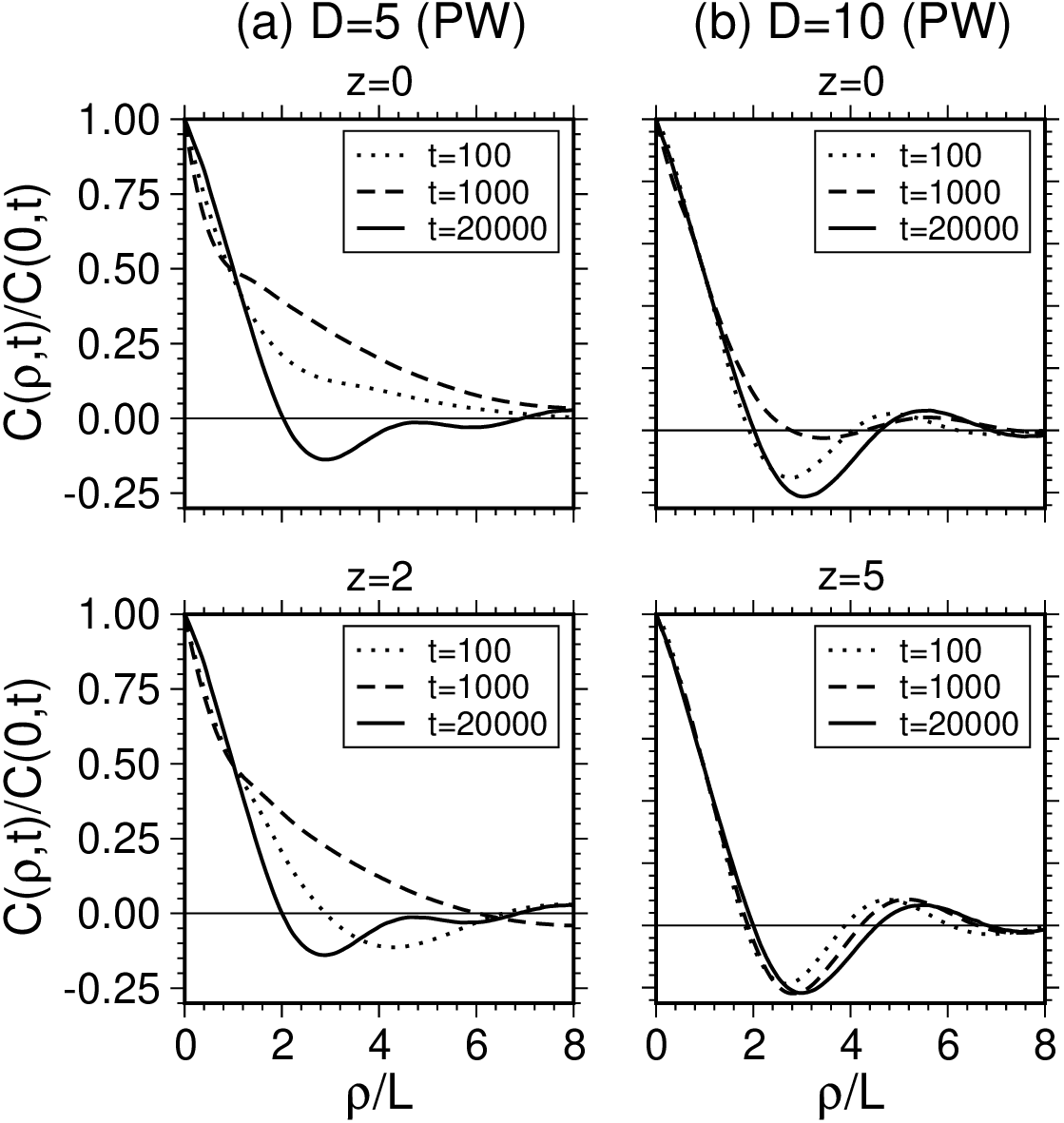}
\caption{\label{fig14}
Scaling plot of layer-wise correlation
functions for the evolution depicted in Fig.~\ref{fig11}. We plot
$C(\rho,t)/C(0,t)$ vs. $\rho/L$ for $t=100,1000,20000$.
We present data for (a) $D=5$ at $z=0$ (wall) and $z=2$ (center);
and (b) $D=10$ at $z=0$ (wall) and $z=5$ (center).}
\end{figure}

\begin{figure}
\includegraphics[width=120mm]{fig15.eps}
\caption{\label{fig15} Time-dependence of the layer-wise
length scale for the evolution depicted in Fig.~\ref{fig11}. We
plot $L(z,t)$ vs. $t$ on a log-log scale for various values of $z$
and (a) $D=5$, and (b) $D=10$.}
\end{figure}

\begin{figure}
\includegraphics[width=110mm]{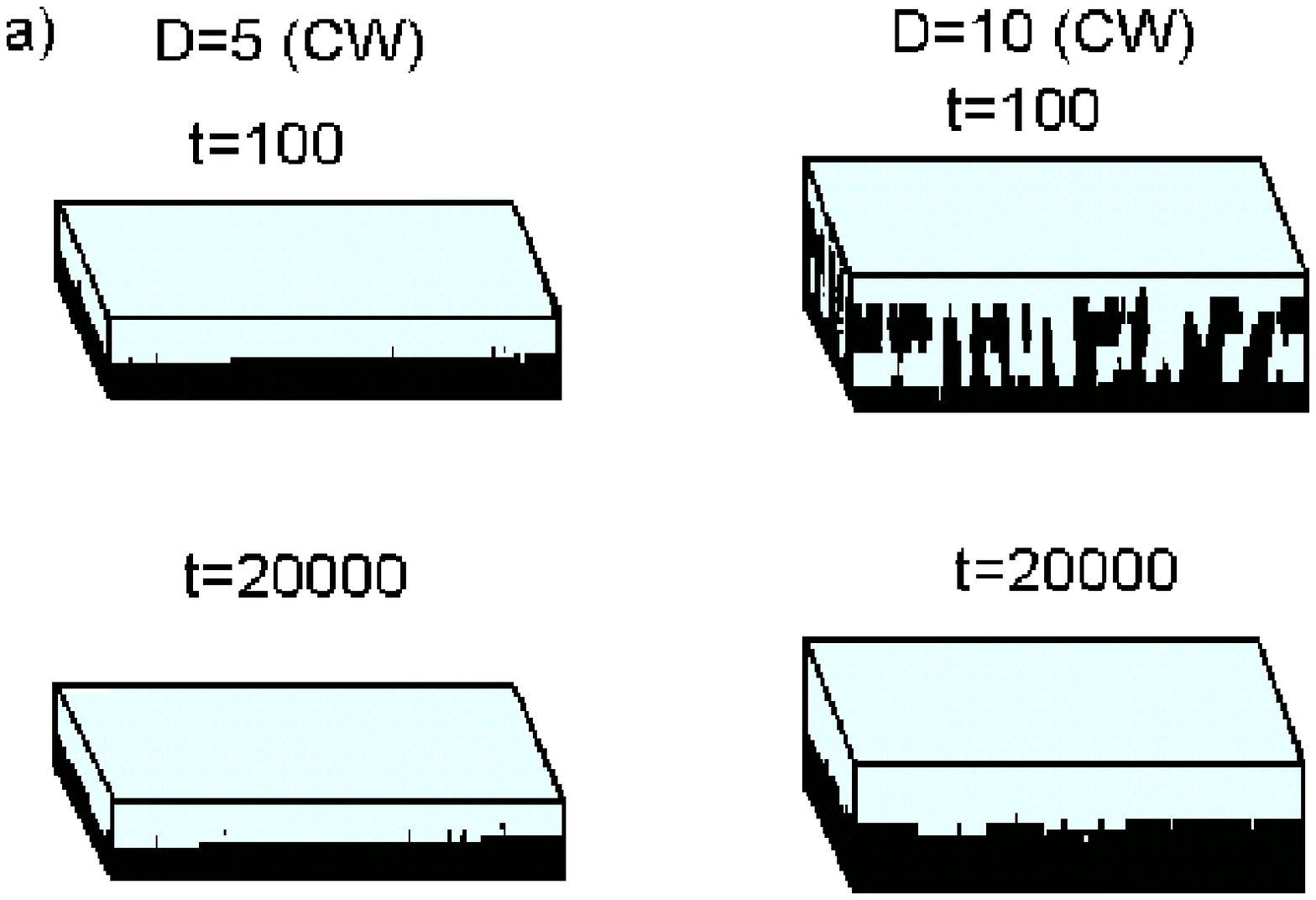}
\includegraphics[width=110mm]{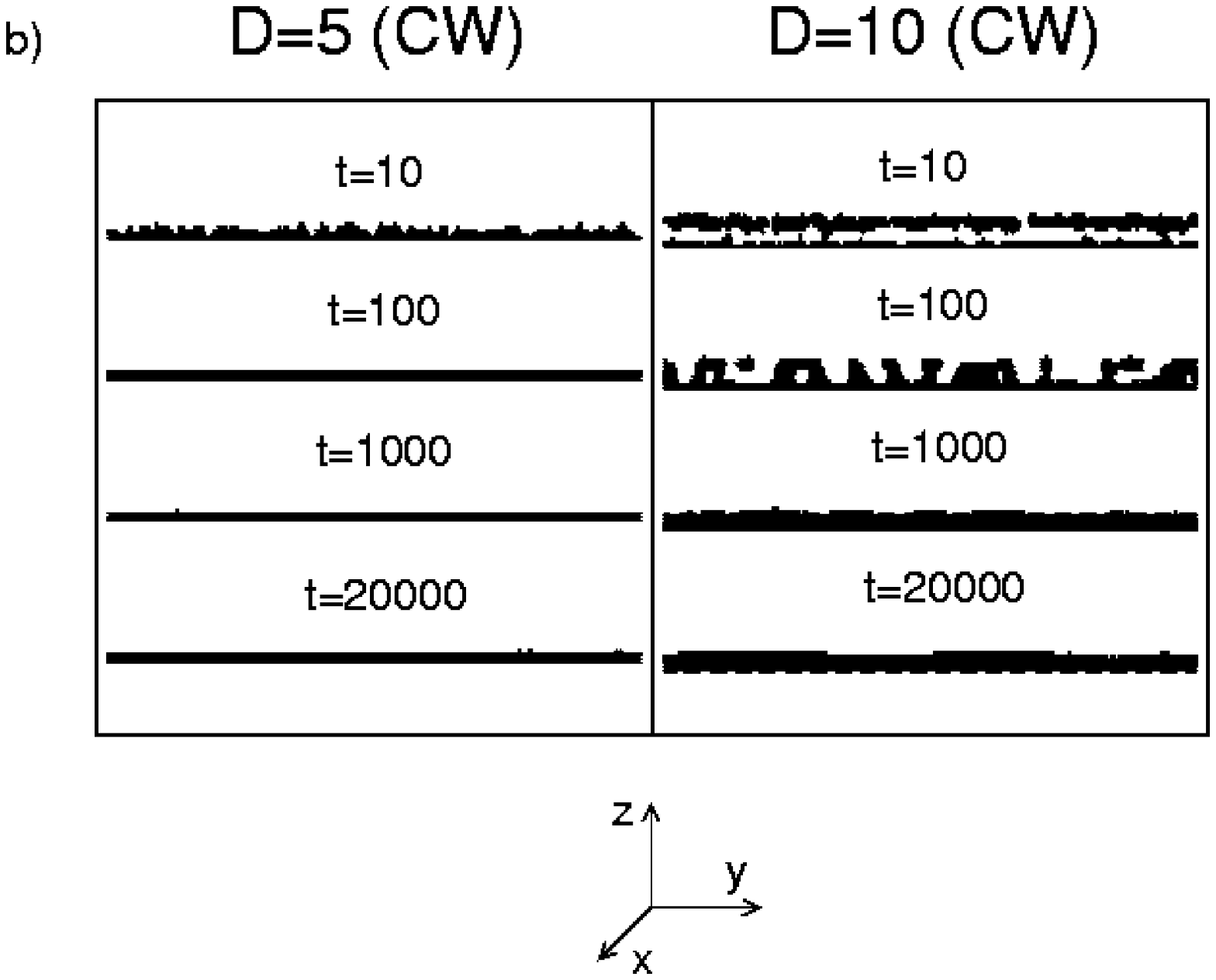}
\caption{\label{fig16}
Analogous to Fig.~\ref{fig2}, but for an antisymmetric film with
a CW morphology.}
\end{figure}

\begin{figure}
\includegraphics[width=120mm]{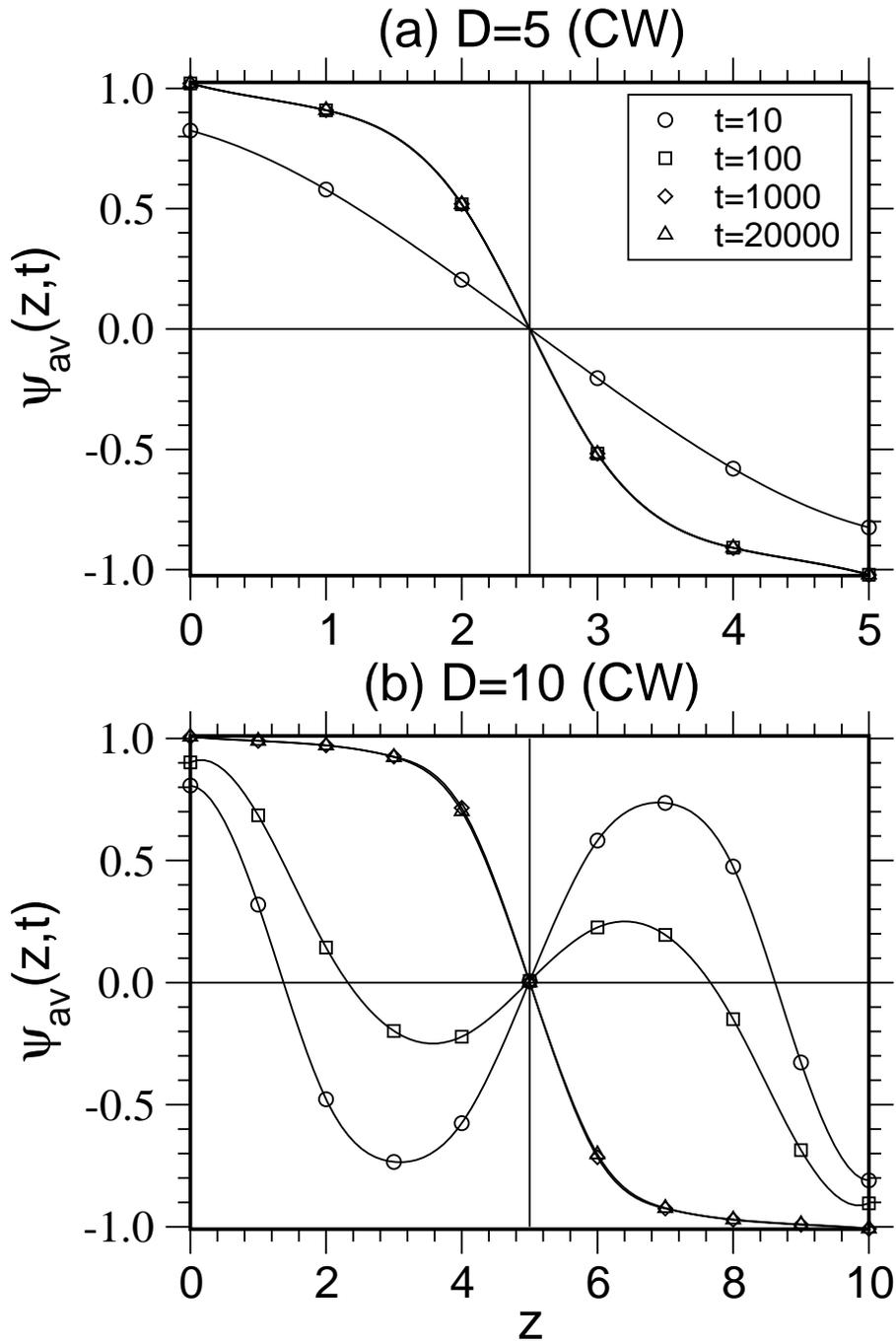}
\caption{\label{fig17} Laterally averaged profiles for the
evolution depicted in Fig.~\ref{fig16}, for (a) $D=5$, and
(b) $D=10$.}
\end{figure}

\begin{figure}
\includegraphics[width=80mm]{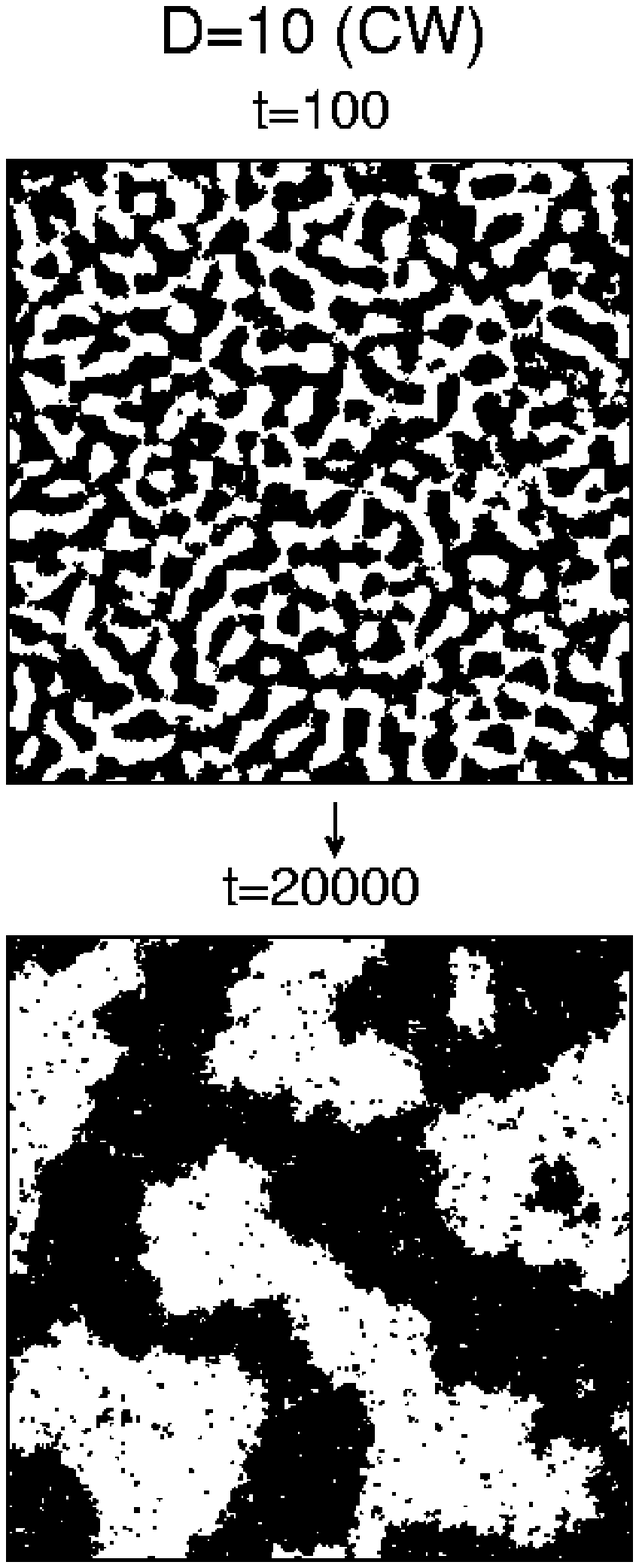}
\caption{\label{fig18}
Cross-sections of the evolution snapshots for $D=10$
in Fig.~\ref{fig16}(a). The cross-section is taken parallel to the
surfaces at $z=5$.}
\end{figure}

\begin{figure}
\includegraphics[width=100mm]{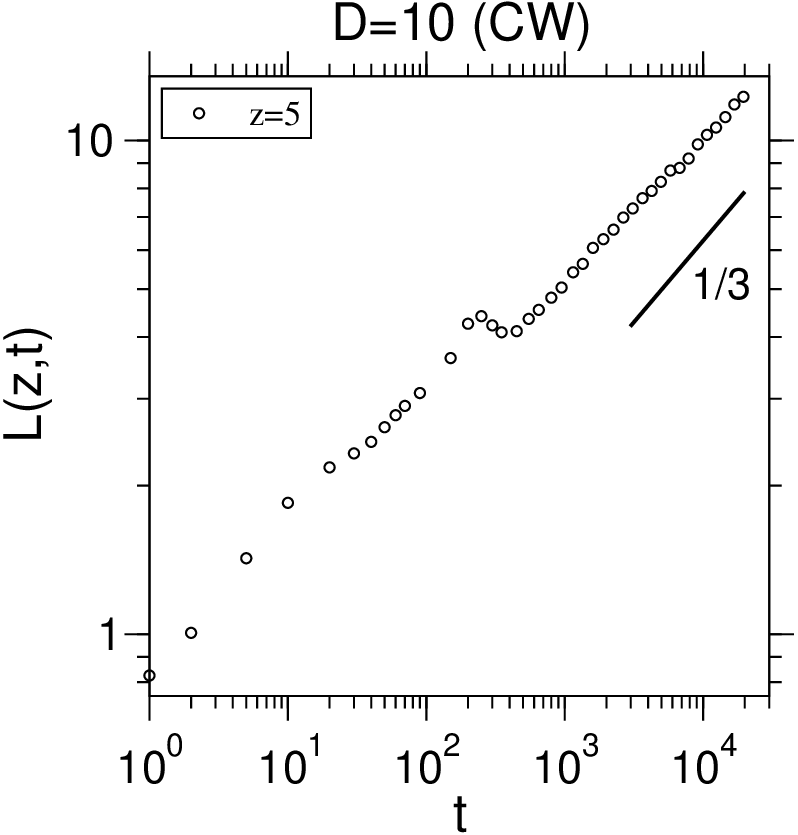}
\caption{\label{fig19} Time-dependence of the layer-wise
length scale for the $D=10$ evolution depicted in Fig.~\ref{fig16}. We
plot $L(z,t)$ vs. $t$ on a log-log scale for $z=5$.}
\end{figure}

\end{document}